\documentclass[twocolumn]{aastex62}

\received{M D, 2019}
\revised{M D, 2019}
\accepted{M D, 2019}
\submitjournal{ApJ}

\shorttitle{Redshift horizon for FIR/sub-mm sources}
\shortauthors{De Rossi \& Bromm}
\begin{document}

\title{Redshift Horizon for Detecting the First Galaxies in Far-Infared Surveys}

\correspondingauthor{Mar\'{\i}a Emilia De Rossi}
\email{mariaemilia.dr@gmail.com}

\author{Mar\'{\i}a Emilia De Rossi}
\affiliation{Universidad de Buenos Aires, Facultad de Ciencias Exactas y Naturales y Ciclo B\'asico Com\'un,
Buenos Aires, Argentina}
\affiliation{CONICET-Universidad de Buenos Aires, Instituto de Astronom\'{\i}a y F\'{\i}sica del Espacio (IAFE),
Buenos Aires, Argentina}

\author{Volker Bromm}
\affiliation{Department of Astronomy, University of Texas at Austin,
2511 Speedway, Austin, TX 78712, USA}
%%\collaboration{(AAS Journals Data Scientists collaboration)}

%\author{Butler Burton}
%\affiliation{National Radio Astronomy Observatory}
%\affiliation{AAS Journals Associate Editor-in-Chief}
%\nocollaboration



\begin{abstract}
%250 word 
We explore the possibility of detecting the first galaxies with the next generation of space-based far infrared (FIR) telescopes by applying an analytical model of primordial dust emission.
Our results indicate that FIR/sub-mm sources at $z\ga 7$ will experience
a strong negative K-correction. Systems of a given virial mass
would exhibit larger dust luminosities at higher $z$, as a consequence of the increase
in dust temperature driven by the higher temperature floor set by the cosmic microwave background.
In addition, high-$z$ systems are more concentrated, which enhances the heating efficiency
associated with stellar radiation. 
By analysing source densities as a function of $z$, and considering 
survey areas of 0.1 ${\rm deg}^2$ and 10 ${\rm deg}^2$, we find that
the redshift horizon for detecting at least one source 
would be above $z\sim 7$ for instrument sensitivities $\la 0.1-0.5 \ {\mu}{\rm Jy}$
and $\la 0.5-3.0 \ {\mu}{\rm Jy}$, respectively, with the exact values depending on
the nature of primordial dust. However, galaxy populations with higher than typical metallicities,
star formation efficiencies and/or dust-to-metal ratios could relax such sensitivity requirements.
In addition, the redshift horizon shows a significant dependence on the nature of primordial dust. 
We conclude that future FIR campaigns could play a crucial role in exploring the nature of dust and 
star formation in the early universe.
\end{abstract}



\section{Introduction}
\label{sec:intro}

Luminous sources at high redshifts ($7 \la z \la 20$), such as the first galaxies and quasars, constitute promising targets 
for next-generation surveys as they provide unique tools to test models of early structure formation \citep{dayal2018}.
According to the standard $\Lambda$-cold dark matter ($\Lambda$CDM) cosmology,
the first stellar systems, the so-called Population~III (Pop~III), formed inside dark matter minihalos of mass 
$\sim 10^6 \ {\rm M_{\sun}}$ at $z \ga 20$ \citep[e.g.,][]{couchman1986, tegmark1997, bromm2004, bromm2009}.
However, the low dynamical (virial) masses of these early cosmic structures render them highly 
susceptible to negative feedback mechanisms that heat the gas, expelling it from their shallow potential wells. Subsequent star formation is thus subject to quenching or delay \citep[e.g.,][]{muratov2013,jeon2014}.
If we define a bona-fide galaxy as a long-lived stellar system, the first protogalaxies in the pre-reionization
universe could have formed inside the deeper potential wells of atomic cooling halos, with virial masses of $\sim 10^7 -  10^8 \ {\rm M_{\sun}}$
at $z \sim 6 - 15$ \citep[][]{oh2002, bromm2011, bromm2011b,safranek2014}.   
In this context, the first supernovae (SNe) would also have engendered the formation of the first dust grains. Detection of thermal dust emission from these primeval sources could thus provide further insight into the nature of primordial 
star formation \citep[e.g.,][]{valiante2009,gall2011,jaacks2018}.

The study of dusty star-forming galaxies (DSFGs) has been an active field of research during the last decade
\citep[see][for a review]{casey2014}. 
The combined population of DSFGs generates an infrared (IR) radiation field equivalent to 
the energy density associated to direct stellar emission from galaxies at 
UV and optical wavelengths.  
Among DSFGs are the most intense starbursts in the universe, reaching star formation rates (SFRs)
of $\sim 1000 \ {\rm M}_{\sun} {\rm yr}^{-1}$ and IR luminosities of $\sim 10^{13} \ {\rm L}_{\sun}$. 
The most luminous of these sources are the brightest galaxies in the universe, and are observed already 
$\sim 800$ Myr after the Big Bang.
However, luminous IR sources constitute a diverse population, ranging from gas-rich disks to galaxy mergers, with 
a sub-set harboring heavily dust-enshrouded supermassive back holes (SMBHs).  
In particular, the discovery of bright sub-millimeter (sub-mm) and IR sources at high-$z$ poses a  
challenge for models of galaxy formation
\citep[e.g.,][]{baugh2005, finlator2006,fontanot2007, dave2010,narayanan2010, somerville2012}, and it is debated whether they are scaled-up versions 
of locally observed systems, such as the ultra-luminous infrared galaxies (ULIRGs), or whether they have a different nature
\citep[e.g.,][]{derossi2018}.

At very high redshifts ($z \ga 7$), reprocessed emission from dust heated by 
stellar UV radiation would be mostly observed at far-IR (FIR) and sub-mm wavelengths.
Only sparse and uncertain observations were reported at these extreme distances, with
few galaxies and active galactic nuclei (AGN) observed at $z \sim 7-8$ \citep[e.g.,][]{mortlock2011, venemans2013}, and 
even fewer at $z\sim 8-11$.
One of the highest redshift sources was detected by \citet{oesch2016} at $z \sim 11$,
which seems to exhibit a stellar mass of $M_{\star} \sim 2 \times 10^9 \ {\rm M}_{\sun}$ 
and an SFR of $\sim 20$ ${\rm M}_{\sun} {\rm yr}^{-1}$.
In addition, dust continuum has also been reported for a few sources at $z > 6$.
\citet{watson2015} detected a significant amount of dust in a $z\approx 7.5$ galaxy,
whose rest-frame UV colors suggested negligible reddening.
More recently, \citet{laporte2017} detected strong mm-continuum flux
from a star forming galaxy during the reionization epoch. This source is located at $z\approx8.4$, hosting
a stellar mass of $M_{\star} \sim 2 \times 10^9 \ {\rm M}_{\sun}$, 
an SFR of $\sim 20 \ {\rm M}_{\sun} {\rm yr}^{-1}$, and a dust mass of $M_{\rm d} \sim 6\times10^6 \ {\rm M}_{\sun}$.
In addition, \citet{venemans2017} analysed the $[{\rm CII}]$ emission and FIR dust continuum in 
one of the most distant quasars known ($z=7.1$), using data from the Atacama Large Millimeter/sub-millimeter Array (ALMA).
These authors inferred a dynamical mass of $\sim 4 \times 10^{10} {\rm \ M}_{\sun}$,
and an SFR of $\sim 100-350$ ${\rm M}_{\sun} \ {\rm yr}^{-1}$ for the host galaxy.

With the advent of new observational facilities to explore the high redshift universe, it is important to provide theoretical models to make predictions about plausible targets, and to guide the development of next-generation instruments.
In this regard, \citet{carilli2017} assessed the prospects for detecting the $[{\rm CII}]$ 158 $\mu$m line 
from distant galaxies, reaching into the `dark ages'. At mm wavelengths, and at 6$\sigma$ significance, ALMA should be able to detect the $[{\rm CII}]$ line emission, integrated over 40\,h, from an actively star-forming galaxy with moderate metallicity out to $z\sim 10$. The next-generation Very Large Array (ngVLA), planned for 2030 and beyond, will have the sensitivity to detect this emission line from a Milky-Way sized galaxy at $z\simeq 15$.  
However, the main challenge is the low surface density of high-redshift sources in blind surveys. Employing results from the {\sc BlueTides} simulation, \citet{wilkins2017} analysed the capabilities of different instruments to study sub-mm sources at $z \ga 8$. They predict that the surface density of $z \ga 8$ sub-mm sources is too low to be detected with {\em Herschel}, SCUBA-2, or ALMA surveys. Nonetheless, improvements in the area coverage of ALMA could achieve the requirements for detection.
The possibility of exploring primeval sub-mm/FIR sources at $z \ga 7$ could be significantly
enhanced with the advent of the next generation of space-based FIR telescopes, such as
the Origins Space Telescope (OST) or the Space Telescope for Cosmology and Astrophysics (SPICA), planned for launch in the 2030s.
Among the key science goals for these facilities will be the cosmic origin of dust, elucidating the first sources of dust emission, which are beyond the current horizon of observability \citep[e.g.,][]{gruppioni2017}.

Indeed, the future space-borne FIR telescopes will be ideally complementary to current (e.g., ALMA) and next-generation facilities, such as the {\it James Webb Space Telescope (JWST)}, which will probe $z\gtrsim 10$ sources in the rest-frame UV. The space-borne missions will be operating in tandem with the suite of extremely large, 30m-class, telescopes on the ground, the Giant Magellan Telescope (GMT), the Thirty-Meter Telescope (TMT), and the European Extremely-Large Telescope (E-ELT). The cold gas at high redshifts will be studied by meter-wavelength radio facilities, focused on the detection of the redshifted $21$~cm radiation from neutral hydrogen
\citep{furlanetto2006, barkana2007}. Among them are the Low Frequency Array (LOFAR), 
the Murchison Wide-Field Array (MWA), 
the Precision Array to Probe the Epoch of Reionization (PAPER), 
the Hydrogen Epoch of Reionization Array (HERA), 
and ultimately the Square Kilometre Array (SKA).

The main aim of this work is to explore the redshift horizon for dust-emitting galaxies, to be reached with a generic next-generation  
FIR telescope, considering different instrument sensitivities and survey areas. We apply an analytical model of primordial dust emission in order to evaluate the possibility of detecting primeval FIR/sub-mm sources at $z \ga 7$,  given our current understanding of star and galaxy formation at early cosmic times. Our study is specifically aimed at dust emission from the first galaxies, which formed after Pop~III stars enriched the ISM with primordial dust grains.  
Typical galaxies at these early epochs ($z\approx7-20$) require special modelling as they are not directly related to typical dusty
galaxies at lower redshifts, which correspond to a later stage of evolution. These systems typically exhibit very low dust densities, 
with FIR/sub-mm fluxes below the capabilities of current and near-future instruments \citep{derossi2017}, rendering their detection very challenging. Future FIR facilities will significantly increase the prospects of achieving the required sensitivity limits. As the nature of the first galaxies is a matter of debate, we explore different dust models and consider systems spanning a wide range of metallicities and dust fractions. The details of such predictions are bound to change, in response to the evolving FIR mission concepts, but it will be useful to establish a baseline exploration.

The plan for the paper is as follows. We discuss our basic theoretical ingredients in Section~2.
In Section~3, we analyse the strong negative K-correction affecting high-$z$ sources, followed by a discussion of the minimum halo mass required for detection in Section~4. Our key results for the redshift horizon for different survey parameters are explored in Section~5. 
In Section~6, we assess the parameter sensitivity of our results. 
We conclude by comparing our predictions for future FIR facilities
with those for the NOrthern Extended Millimeter Array (NOEMA), ALMA (Section~7) and the {\it JWST} (Section~8). 
A brief summary is presented in Section~9. The following cosmological parameters are assumed in this work:
$h$ = 0.67,
${\Omega}_{\rm b}$ = 0.049,
${\Omega}_{\rm M}$ = 0.32,
${\Omega}_{\Lambda}$ = 0.68
\citep{planck2014}.

\begin{figure}
\plotone{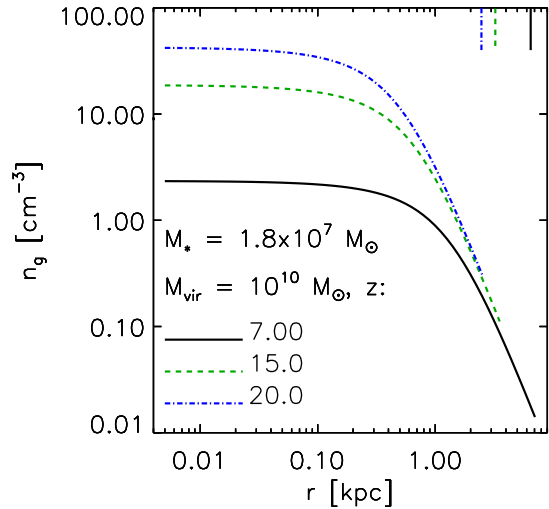}
	\caption{Density structure within the first galaxies. Shown is gas number density as a function of radius, for systems of
	$M_{\rm vir} = 10^{10} \ {\rm M}_{\sun}$ and stellar mass $M_* \approx 1.8\times10^7\ {\rm M}_{\sun}$ at different $z$, as indicated.
The vertical lines on the top depict the corresponding virial radii. Evidently, galaxies are more compact at higher $z$, with important consequences for their intrinsic emissivities.
}
    \label{fig:ng_vs_r}
\end{figure}

\section{Methodology}
\label{sec:methodology}

To explore the detectability of FIR/sub-mm sources at $z \ga 7$, 
we apply the model for dust emission from primeval galaxies developed in
\citet{derossi2017}.
For the convenience of the reader, the model is briefly described
in Sections~\ref{sec:first_galaxies_model} and~\ref{sec:dust_model} (for more details, see
\citealt{derossi2017}).
In Section~\ref{sec:density_model}, we introduce the methods used to calculate source densities at different redshifts, allowing us to derive the redshift horizon as a function of instrument sensitivity, given our dust model.

\begin{figure*}
	\begin{center}
\plottwo{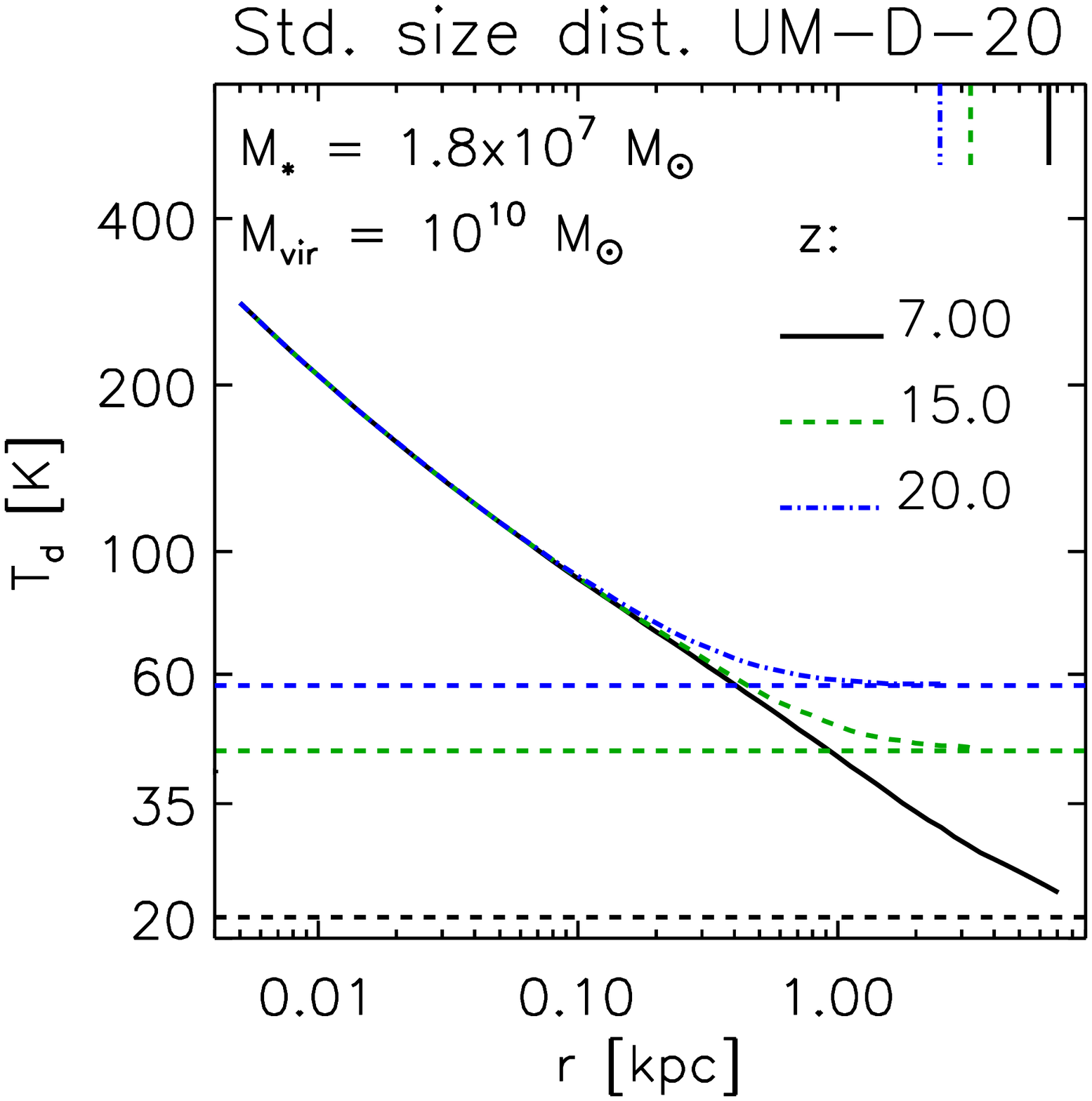}{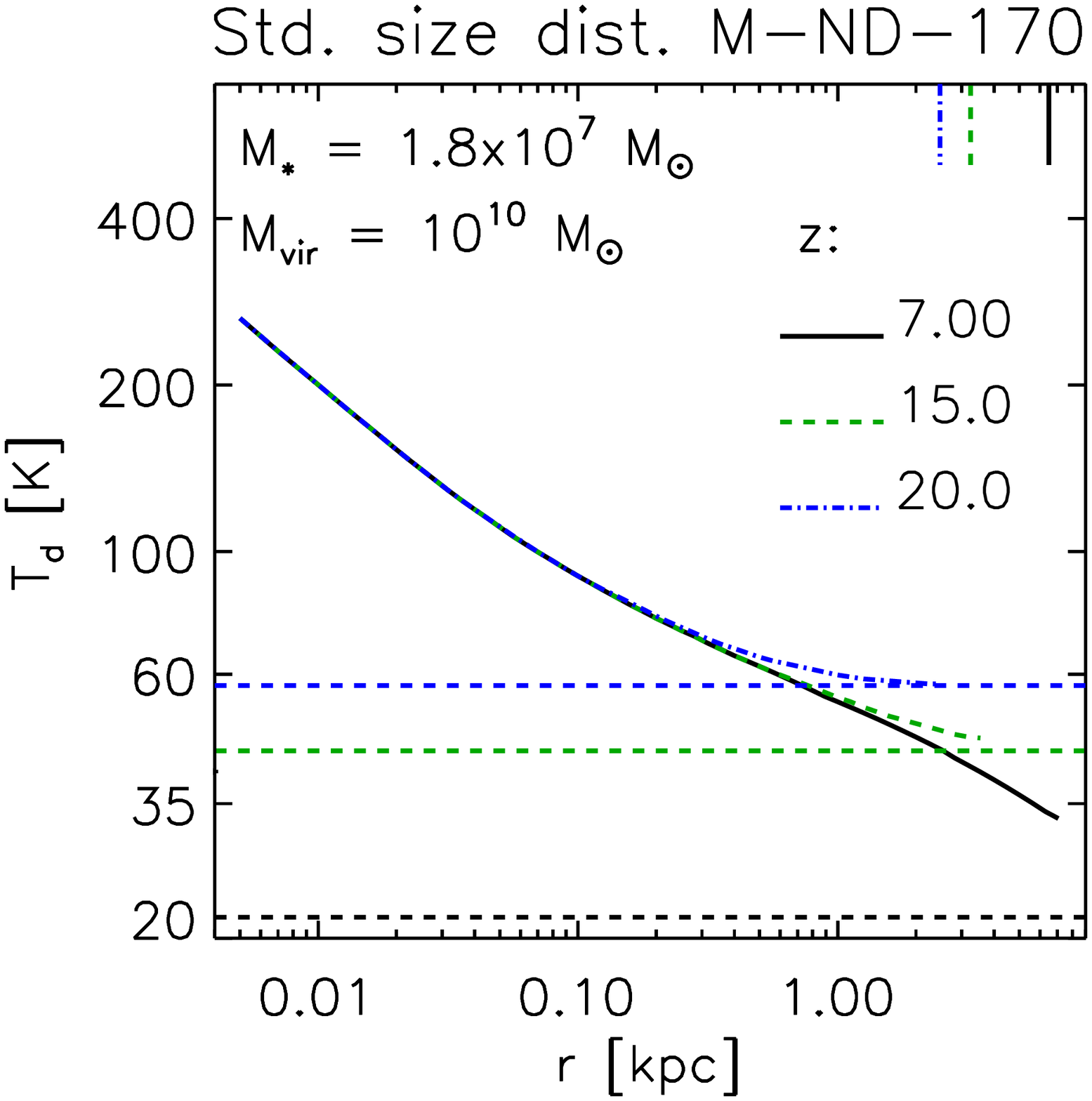}
	\end{center}
    \caption{Dust temperature profiles for systems similar to those shown in
Fig. \ref{fig:ng_vs_r}. The standard size distribution for dust grains is implemented
using two different dust chemical compositions: UM-D-20 (left-hand panel) and M-ND-170 (right-hand panel).
The horizontal dashed lines denote the CMB temperature at the different redshifts considered.
The vertical lines on the top depict the virial radii of these systems.}
    \label{fig:Td_vs_r}
\end{figure*}

\subsection{Modelling first galaxies}
\label{sec:first_galaxies_model}

For modelling a first galaxy, we consider a virialized dark matter halo hosting a central metal-poor Population~II (Pop~II) stellar cluster,
surrounded by a mixed phase of gas and dust.
We make the idealized assumption of spherical symmetry and adopt a gas-density profile 
of the \citet{burkert1995} form\footnote{Other density profiles, such as isothermal 
or Navarro-Frenk-White (NFW), lead to similar results \citep[see][]{derossi2017}.}.
Following \citet[][]{derossi2017}, the gas profile is defined in such a way
that the baryon-to-total mass ratio of a given galaxy is of the order
of the cosmic mean (${\Omega}_{\rm b} / {\Omega}_{\rm M}$).  
For simplicity, our model does not consider stellar feedback effects,  
which could expel part of the interstellar medium (ISM) out of the systems; hence, the derived dust 
luminosities constitute upper limits.
In Fig. \ref{fig:ng_vs_r}, we show gas density profiles for model
galaxies with a virial mass of $M_{\rm vir} \sim 10^{10} \ {\rm M}_{\sun}$
at select redshifts. The vertical lines on the top indicate the corresponding
virial radius, $R_{\rm vir}$, for each system. It is evident that systems at
higher $z$ are more concentrated. In particular, central gas densities are higher
by more than one order of magnitude at $z\sim20$, compared to $z\sim7$.

To assign the mass of the stellar
component, we assume a star formation efficiency of $\eta = M_* / (M_{\rm g} + M_*) = 0.01$,
where $M_*$ and $M_{\rm g}$ represent the stellar and gas mass, respectively
\citep[e.g.,][]{greif2006, mitchellwynne2015}.
Considering that the baryon-to-total mass ratio of our primeval sources is 
of the order ${\Omega}_{\rm b}/{\Omega}_{\rm M}$, the stellar-to-total mass ratio
($M_* / M_{\rm vir}$)
for these systems is about $\approx 2\times10^{-3}$.
Such $M_* / M_{\rm vir}$ corresponds to
the minimum value usually derived from models that parametrize the
relation between galaxy and halo assembly \citep[e.g.,][]{behroozi2019}.
Based on the behavior observed at lower redshifts, extrapolating these models to
$z>7$ and $M_{\rm vir} \la 10^{13}\ {\rm M}_{\sun}$, give $M_* / M_{\rm vir}$ 
ranging between $\sim 0.001 -0.1$, with the exact value depending on model details, mass and
$z$. However, given that the estimates of $M_*$, SFR, and dust content at $z>4$ remain quite uncertain,
the behavior of such quantities at extremely high-$z$ is still vigorously debated.
Here, as we are modelling the very {\em first} galaxies (formed after Pop III stars enriched the ISM
with the first metals and dust), we adopt a conservative $\eta =  0.01$ (see 
Section~\ref{sec:parameter_variations} for results with a higher $\eta$).

The spectral energy distribution associated with stars is derived by using Yggdrasil model
grids \citep{zackrisson2011}, corresponding to the lowest available stellar metallicity,
$Z_* \approx 3 \times 10^{-2} Z_{\sun}$, and a stellar age $\tau = 0.01 {\rm \,Myr}$.
To derive dust densities, we adopt a dust-to-metal mass ratio of
$D/M = M_{\rm d} / M_{\rm Z} = 5 \times 10^{-3}$, and a gas metallicity of
$Z_{\rm g} = 5 \times 10^{-3} Z_{\sun}$. Note that the dust-to-metal ratio here is a global ratio of masses in the system, different from the local ISM density-ratio often considered, with the local ratio being much larger. Our choice of parameters results in a total dust mass that follows the scaling observed in I~Zw~18, which is a local analogue of extremely metal-poor galaxies at high redshifts \citep[e.g.,][]{schneider2016}.

\subsection{High-redshift dust emission}
\label{sec:dust_model}

Following \citet{ji2014}, we consider a suite of different silicon-based dust
models, which were calculated by \citet{cherchneff2010} analysing the ejecta of Pop III 
supernovae:
UM-ND-20, UM-ND-170,
UM-D-20, UM-D-170, M-ND-20, M-ND-170, M-D-20 and M-D-170.\footnote{Model names follow 
the same notation used in \citet{cherchneff2010}:
UM = unmixed, M = mixed; ND = non depleted, D = depleted; 170: 170 ${\rm M}_{\sun}$ progenitor,
20: 20 ${\rm M}_{\sun}$ progenitor.}
We acknowledge that there is an ongoing debate to what extent dust at early cosmic times contains a carbon-based component in addition to the silicon-based one.
\citet{cherchneff2010} implemented non-equilibrium
chemical kinetics for modelling dust formation.  In these models,  the suppression of carbon dust 
relies on the hypothesis that carbon-rich regions in the supernova ejecta are microscopically 
mixed with helium ions. On the other hand, most steady-state models would predict a dominant carbon 
dust component.
It is worth emphasizing that we here focus on
the primeval galaxy stage ($< 400$ Myr since the beginning of star formation), during which 
stars less massive than 3.5 ${\rm M}_{\sun}$ should not have ejected significant amounts of carbon-enriched
material into the ISM.\footnote{A similar approach is adopted in \citet{derossi2017} and \citet{derossi2018}.}
As discussed in \citet{derossi2017}, if a moderate contribution of carbon were considered in our model, 
we would expect slight variations in the dust temperature and enhanced dust emissivities by 
a factor of a few.

\citet{cherchneff2010} do not provide grain size distributions for their models. Thus, for 
modelling the grain size distribution, we consider the so-called ``standard'' and ``shock'' prescriptions adopted by \citet{ji2014}. 
The standard distribution \citep{pollack1994} is similar to the Milky Way grain size distribution, 
while the shock distribution is based on the work of \citet{bianchi2007}, approximating the effect 
of running a post-explosion reverse shock through newly created dust, which leads to smaller
grains. Different types of dust would follow likely different size distributions as 
different chemical species condense to different initial sizes and are subject to different amounts of
destruction in a supernova reverse shock (e.g.,
\citealt{todini2001,bianchi2007,nozawa2007,silvia2010}).  However, as the nature of primordial dust
is quite uncertain, for the sake of simplicity, 
we follow \citet{ji2014} and assume a similar size distribution for all our dust species.
Our aim is not to determine which dust model is more realistic, but to explore the sensitivity of 
our results on different dust chemistries and grain size distributions.
Finally, we note that our model does not consider dust formation in the ISM, as this process is likely
to be delayed past the timescale that we consider here, before a significant contribution of carbon dust is injected into the ISM by low mass stars \citep{asano2011}.

In order to estimate the dust temperature ($T_{\rm d}$), we assume thermal equilibrium between the dust cooling and heating rates, the latter given by gas-dust collisions and stellar radiation. As shown by \citet{derossi2017}, photo-heating is the dominant heating mechanism, because of the low ISM densities encountered in the primeval galaxies. The cosmic microwave background (CMB) determines the temperature floor ($T_{\rm CMB}$), as it is thermodynamically not possible to radiatively cool below it.
In Fig. \ref{fig:Td_vs_r}, we show $T_{\rm d}$ profiles for systems similar to those in
Fig.  \ref{fig:ng_vs_r}, but assuming different dust chemical compositions (UM-D-20 and M-ND-170).
We adopt a standard size distribution for dust grains; had we used the shock size
distribution instead, similar trends would have been obtained, but with slightly higher $T_{\rm d}$
values \citep{derossi2017}.  In Fig. \ref{fig:Td_vs_r}, we see that, for a given $M_{\rm vir}$ and 
dust model, systems reach higher dust temperatures ($T_{\rm d}$) at higher $z$ due to the
higher values of $T_{\rm CMB}$.  Also, at
higher $z$, $R_{\rm vir}$ decreases and the density increases towards the center (Fig.  \ref{fig:ng_vs_r}).
Thus, the fraction of dust near the central stellar cluster increases and the
heating is more efficient at higher $z$. Therefore, for systems of similar masses,
we expect larger dust luminosities at higher $z$, as we will confirm in Section \ref{sec:k_correction}.

In order to estimate dust emissivities ($j_{\nu}$) from the $T_{\rm d}$ profiles, 
Kirchhoff's law was applied:

\begin{equation}
j_{\nu} ( T_{\rm d} ) = 4 \pi {\kappa}_{\nu} B_{\nu} (T_{\rm d}),
\end{equation}

where ${\kappa}_{\nu}$ is the frequency-dependent dust opacity per unit mass,
and $B_{\nu}$ is the Planck function.

We estimate the total specific {\em dust} luminosity, $L_{\nu ,{\rm em}}$, emitted
by a given model galaxy, by integrating $j_{\nu}$ out to $R_{\rm vir}$.
Thus, for a galaxy of a given $M_{\rm vir}$ at redshift $z$, the observed {\em dust} 
specific flux $f_{\nu , {\rm obs}}$ originating from the model source is given by:
\begin{equation}
\label{eq:flux}
f_{\nu , {\rm obs}} = (1 + z)  \frac{L_{\nu ,{\rm em}}}{4 \pi {d_{L}}^2}
\mbox{\, ,}
\end{equation}
where $d_{L}$ is the luminosity distance to a source at redshift $z$. 
%(the reader is referred to \citet{derossi2017}, for more details regarding this dust model).
As explained below, in order to analyse the possibility of detecting a system of a given $M_{\rm vir}$
at redshift $z$, we calculate the average $f_{\nu , {\rm obs}}$ over the broad FIR band
$\Delta \lambda = 250-750 \ {\mu}{\rm m}$ and compare it with a potential range of detector sensitivities
at similar wavelengths.

It is worth mentioning that our dust model has been successfully applied to predict
the FIR spectral energy distribution (SED) of first massive galaxies at
$z \gtrsim 5$ \citep{derossi2018}.  In particular, adapting the model to conditions in FIR galaxies
at $z\sim6$ and adding a low percentage of carbon dust, 
\citet{derossi2018} were able to reproduce the SED of Haro 11, which is
a local galaxy with a moderately-low metallicity, an exceptionally high star formation rate 
and a very young stellar population,
which likely approximates the relevant conditions in young massive Pop II galaxies at high $z$.
These authors conclude that there seems to be a progression with redshift in 
observed galaxy SEDs, from those resembling local ones at $1< z <4$ to a closer
resemblance to Haro 11 at $5 \lesssim z < 7$.

Typical rest-frame and observed spectra of model galaxies at $z \gtrsim 7$ in the context of potential 
detections with next-generation FIR facilities are presented and discussed in Section~\ref{sec:k_correction}.

\begin{figure*}
	\gridline{\fig{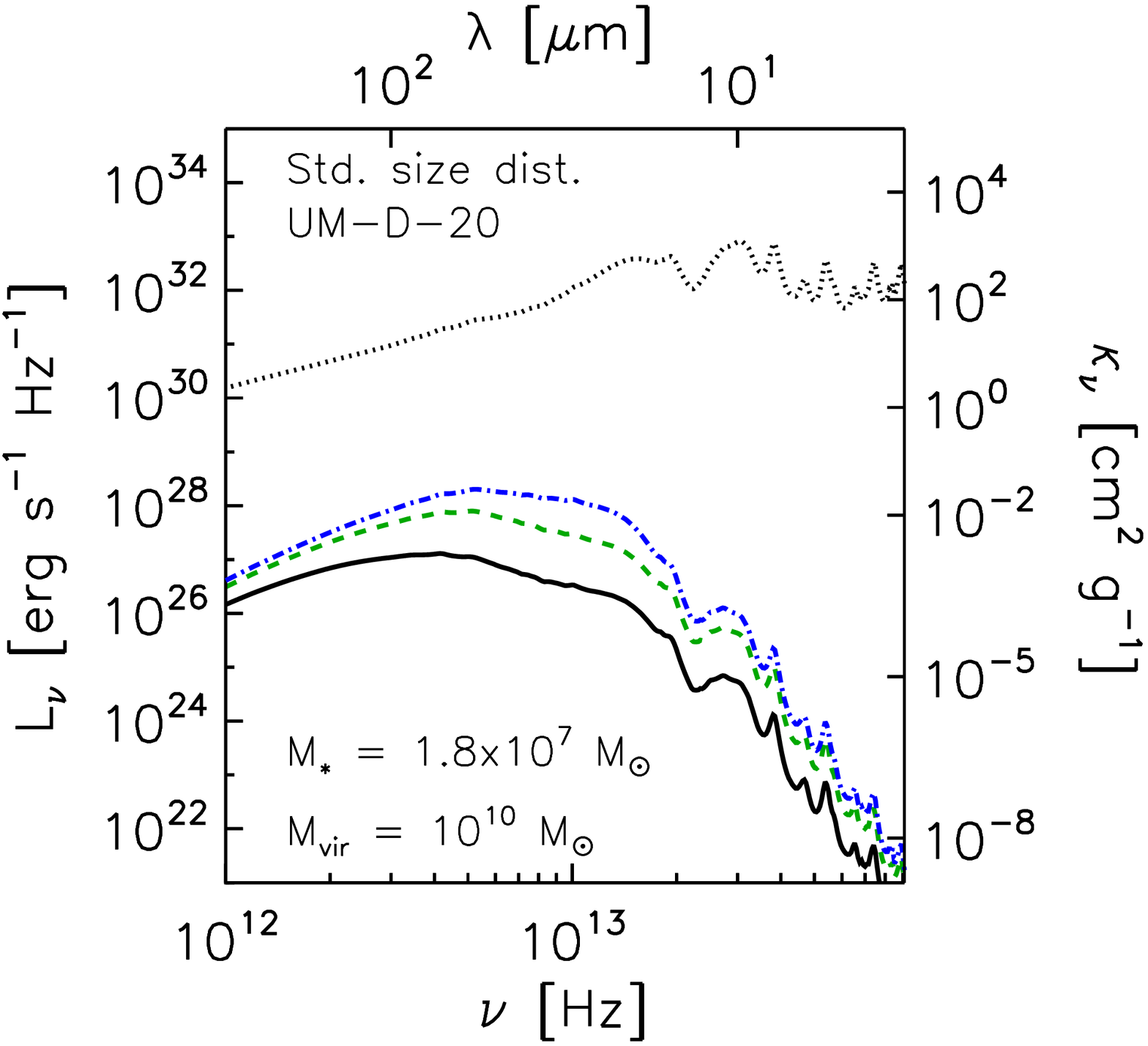}{0.45\textwidth}{(a)}
          \fig{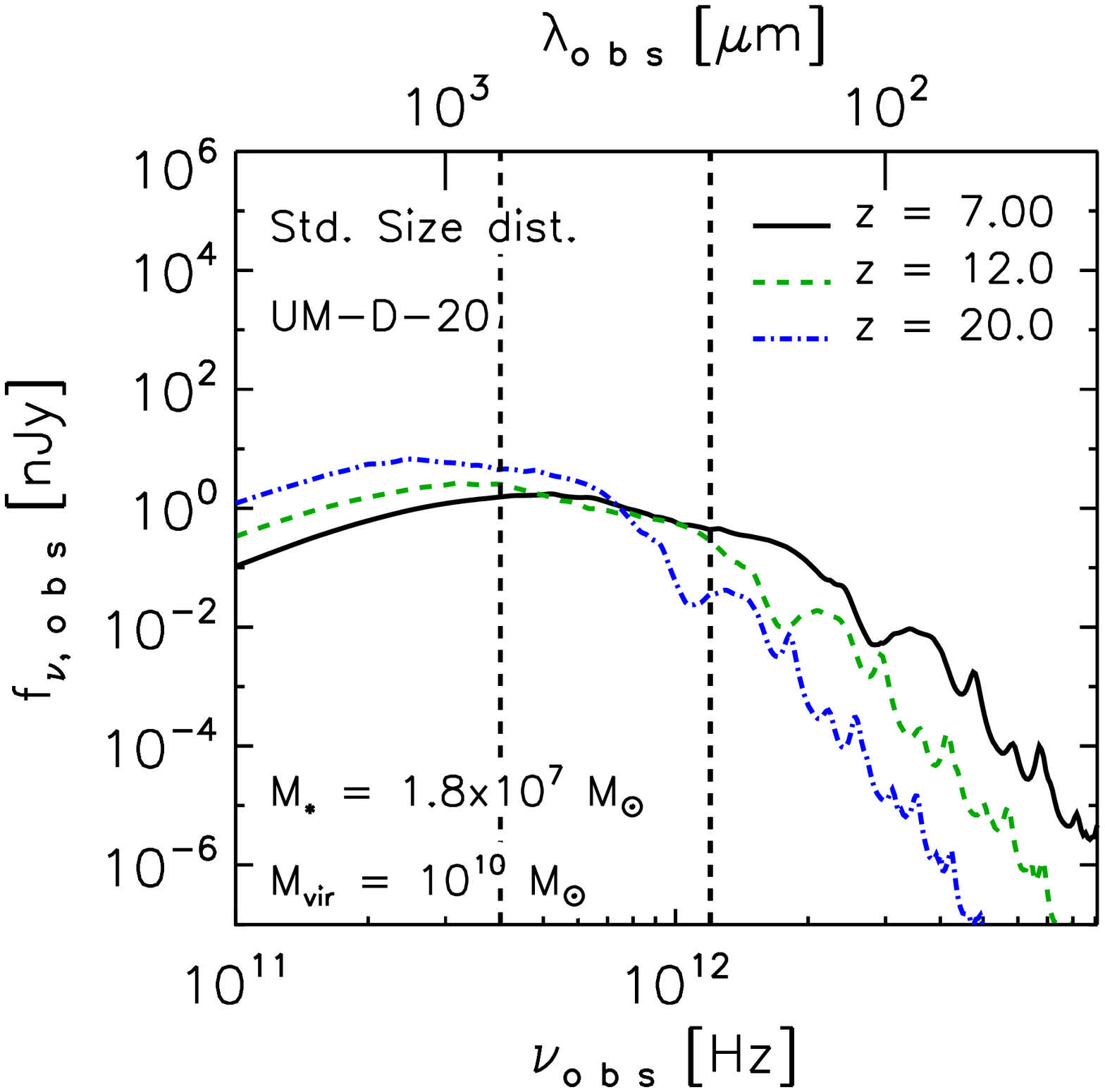}{0.45\textwidth}{(b)}
          }
\gridline{\fig{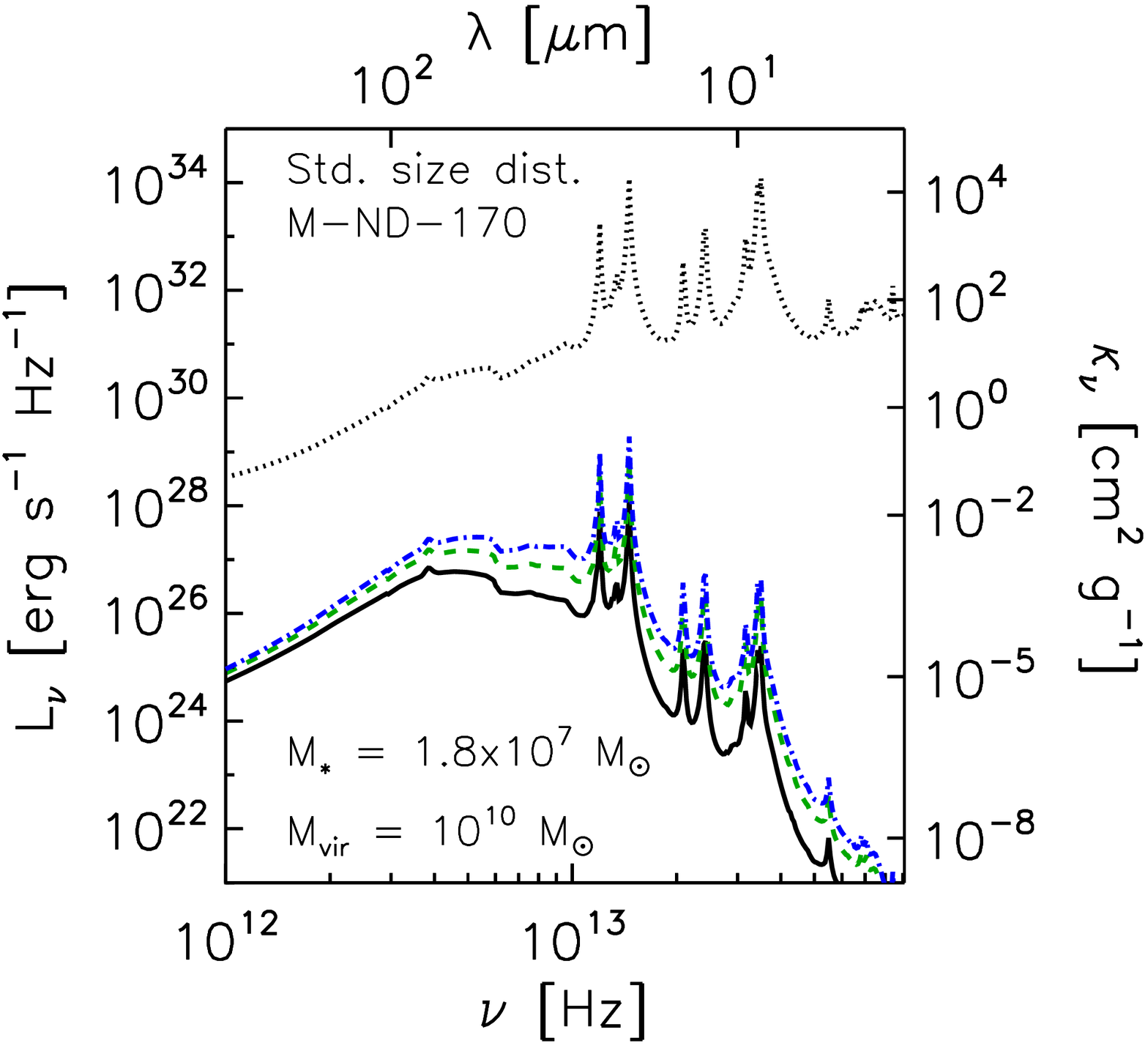}{0.45\textwidth}{(d)}
          \fig{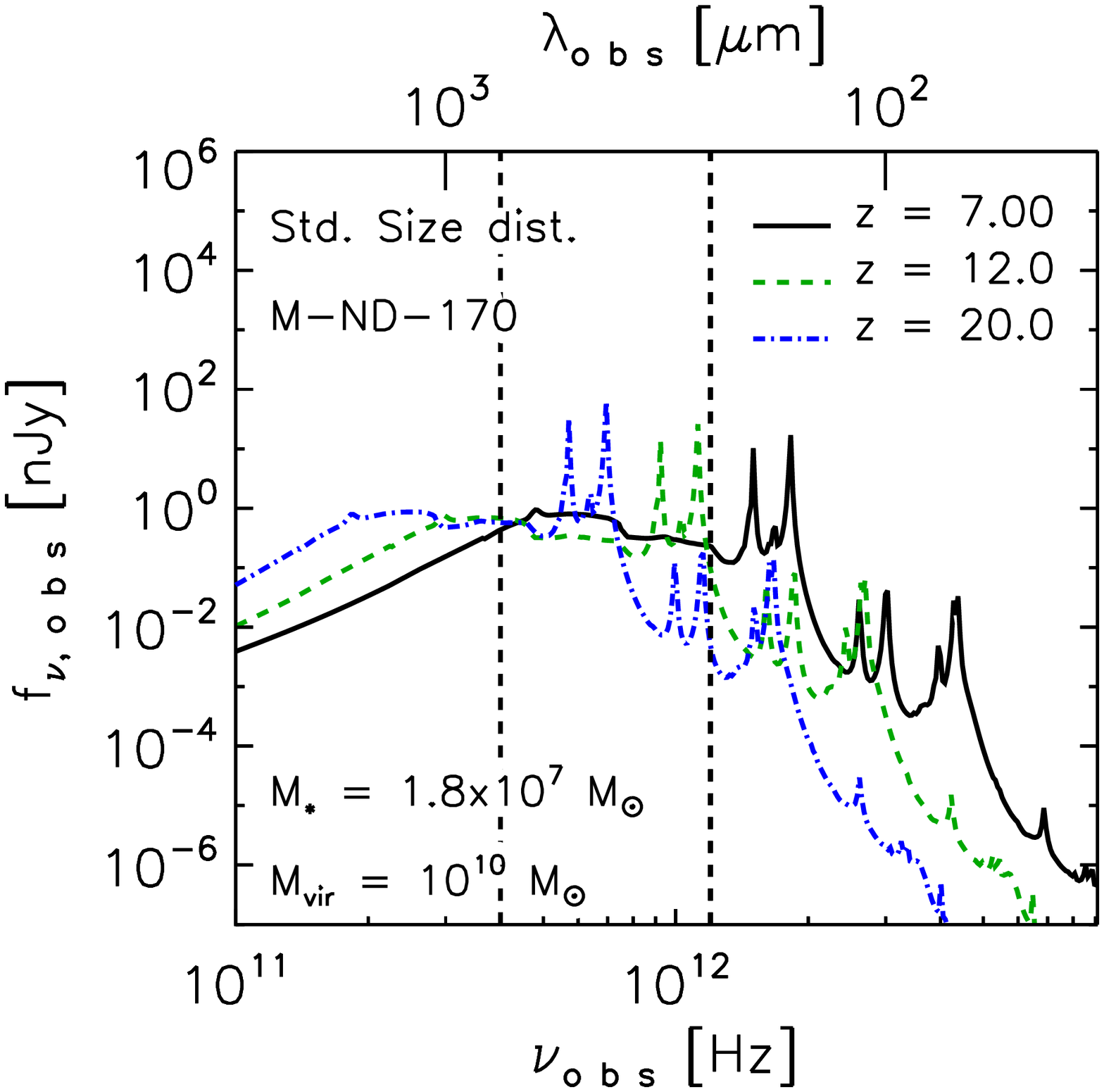}{0.45\textwidth}{(e)}
          }
    \caption{
	    Dust re-emission spectra for the same systems shown in Fig. \ref{fig:Td_vs_r}, all having $M_{\rm vir}=10^{10}\ {\rm M}_{\sun}$ and $M_{\rm *}\approx1.8\times 10^{7}\ {\rm M}_{\sun}$, at different $z$. Left-hand panels:
rest-frame spectra for individual sources. The dotted black curves depict the 
frequency-dependent dust opacities, ${\kappa}_{\nu}$, corresponding to the different dust models.
	Right-hand panels: observed specific fluxes. Vertical dashed lines indicate the FIR frequency range which was adopted in this work.
}
    \label{fig:Lnu_fnu_vs_nu}
\end{figure*}

\subsection{Source densities}
\label{sec:density_model}

Our goal is to estimate the redshift horizon for 
a generic FIR telescope, considering the sensitivity limits required to
detect the weak continuum radiation from primordial dust,
in the context of our current understanding of first galaxy formation.
The redshift horizon ($z_{\rm lim}$)
is defined as the highest $z$ above which the projected number of sources
per given solid angle ($\Delta \Omega$)  
is $N \le N_{\rm crit}$, where we here
consider $N_{\rm crit} = 1$.

For simplicity, we assume that the observed number of sources per unit redshift 
per unit solid angle is given by

\begin{equation}
\frac{dN}{dzd\Omega} =  
\int_{M_{\rm min}}^{\infty} \frac{dN}{dMdV} \frac{dV}{dzd\Omega} dM = 
\int_{M_{\rm min}}^{\infty} n_{\rm ST}\ r^2 \frac{dr}{dz} dM  \mbox{\, ,} 
\end{equation}

\noindent
where $dV$ is the comoving volume element, and $n_{\rm ST}$ the 
Sheth-Tormen mass function\footnote{The 
Sheth-Tormen formalism provides a better fit to simulations
than the Press-Schechter (PS) approach, which 
underpredicts the abundance of halos at high redshifts.} \citep{sheth2001}. The comoving distance to redshift $z$ is given by

\begin{equation}
r(z) = \frac{c}{H_0} \int_{0}^{z} \frac{dz'}{\sqrt{{\Omega}_{\rm m}
(1+z')^3 + {\Omega}_{\Lambda}}} \mbox{\, ,}
\end{equation}

\noindent
where $c/H_0$ is the present-day Hubble distance.

For a given instrument sensitivity ($S$), $M_{\rm min}$ is defined as the virial mass 
corresponding to the least massive systems which could be detected at FIR/sub-mm 
wavelengths at redshift $z$.
To estimate $M_{\rm min}$, we calculate the FIR/sub-mm observed fluxes for
our model galaxies as a function of mass and $z$ (see Section~\ref{sec:dust_model}, 
Equ.~\ref{eq:flux}), and 
compare them with the detector sensitivity at similar wavelengths.
To obtain an estimate of the potential observed FIR/sub-mm fluxes ($F_{\rm FIR}$),
we average the model spectra over the broad FIR band $\Delta \lambda = 250-750 \ {\mu}{\rm m}$, where we approximate
$\Delta \lambda \sim \lambda$:

\begin{equation}
\label{eq:average_flux}
F_{\rm FIR} = \frac{\int_{{\nu}_{i}}^{{\nu}_{f}} f_{\nu , {\rm obs}} \ {\rm d}{{\nu}_{\rm obs}}}{{{\nu}_{f}}-{{\nu}_{i}}}.
\end{equation}

${{\nu}_{i}}$ and ${{\nu}_{f}}$ correspond to the frequency range associated to 
our adopted FIR filter.
To render our results applicable for different sets of instrument
parameters, we analyse results for a wide range of sensitivities
($S=10^{-3} - 10^{3} \ {\mu}{\rm Jy}$), but we note that a sensitivity of $\sim 1 \ {\mu}{\rm Jy}$
would be a more realistic value for the next decades.
Therefore, $M_{\rm min}$ is estimated as the lowest $M_{\rm vir}$ such that
$F_{\rm FIR} \ge S$, where $S\equiv S(t)$ represents the detector sensitivity for a given
exposure time $t$.
As $F_{\rm FIR}$ depends on the adopted dust model, $M_{\rm min}$ and the source density
will also be affected by the adopted dust prescriptions.  

The projected number of sources within a solid angle $\Delta \Omega$ at redshift $z$ is defined as

\begin{equation}
\label{eq:N_z}
N(>z) = \Delta \Omega \int_{z}^{z_{\rm max}} \frac{dN}{dz'd\Omega} dz' \mbox{\, ,}
\end{equation}

\noindent
where $\Delta \Omega$ represents the observed region of the sky, and $z_{\rm max}$ corresponds
to the maximum $z$ where the first galaxies could have formed.
Following \citet{derossi2017}, we adopt $z_{\rm max} = 20$.
We will, particularly, focus on the cases $\Delta \Omega = 0.1, 1.0$ and $10 \ {\rm deg}^2$, 
which correspond to potential ultra-deep, deep and medium survey areas in the FIR.

We emphasize that our adopted sensitivities and survey areas cover a wide range of plausible
values in order to be useful for different potential FIR observational campaigns.

\section{First Galaxy K-Correction}
\label{sec:k_correction}

In this section, we implement the methodology described in Section~\ref{sec:methodology}
to discuss the strong negative K-correction, affecting dust emission
at extremely high $z$.  We show that such K-correction effects significantly enhance
the capabilities of a generic FIR telescope
to detect the dust continuum emission from primeval galaxies at $z=7-20$.

\begin{figure}
\plotone{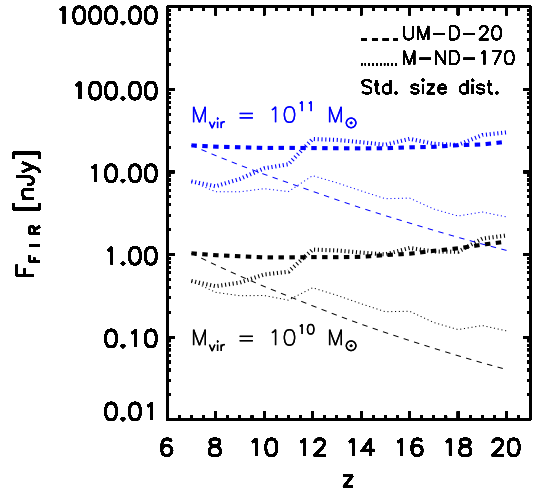}
        \caption{
		FIR flux for galaxies of different masses ($M_{\rm vir} = 10^{10}$ and $10^{11} \ 
	{\rm M}_{\sun}$, corresponding to $M_{\rm *} \approx 1.8\times 10^{7}$ and $1.8\times10^{8} \ {\rm M}_{\sun}$, respectively)  
		as a function of redshift (thick lines).
Results correspond to the same dust models analysed in Fig. \ref{fig:Lnu_fnu_vs_nu}.
In order to show the effects of the strong negative K-correction affecting these galaxies,
thin lines depict the FIR flux obtained assuming that galaxies with similar
masses exhibit the same ($z=7$) rest-frame specific luminosity at all redshifts.
}
    \label{fig:Ffir_vs_z}
\end{figure}

In Fig. \ref{fig:Lnu_fnu_vs_nu}, we show the dust re-emission spectra for the same systems 
shown in Fig.~\ref{fig:Td_vs_r}, all with $M_{\rm vir}=10^{10}\ {\rm M}_{\sun}$ and 
$M_{\rm *} \approx 1.8\times 10^{7}$, at different $z$. 
The rest-frame spectra for individual sources are plotted in the left-hand panels. 
The dotted black curves depict the frequency-dependent dust opacities, ${\kappa}_{\nu}$, 
corresponding to the different dust models.
The distinct spectral features reflect the properties of the underlying opacity curves.
Interestingly, at a fixed $M_{\rm vir}=10^{10}\ {\rm M}_{\sun}$, the maximum luminosity increases
with $z$, with galaxies being $\sim 1$ order of magnitude brighter at $z\approx20$ compared
to similar galaxies at $z\approx7$.
This behaviour is consistent with the $T_{\rm d}$ profiles shown in Fig. \ref{fig:Td_vs_r}.
As we anticipated, higher dust temperatures are expected at high redshifts due to the
higher intensity of the CMB. Also, systems of similar $M_{\rm vir}$
exhibit smaller $R_{\rm vir }$ at higher $z$. Thus, a larger fraction of
dust is located in the surroundings of the central stellar cluster at higher $z$, leading
to a more efficient heating of dust grains. All these effects result in an increased dust emissivity
with increasing $z$. Thus, our findings suggest that the first galaxies would
experience a strong negative K-correction, thus facilitating their
detectability.

In the right-hand panel of Fig.~\ref{fig:Lnu_fnu_vs_nu}, we show observed specific
fluxes, predicted for the two dust models described above. 
The vertical dashed black lines depict the FIR wavelength band adopted in this work
to estimate the average FIR/sub-mm radiation (see Section~\ref{sec:methodology}). As can be seen, 
at a given mass, the observed flux
reaches higher values at higher $z$, despite the larger luminosity distance to the observer. As explained before, the higher
dust temperature achieved at higher $z$ compensates for the distance effect.
We also note that main spectral features enter the FIR band in the $z\sim 12-20$ range.
Thus, observations in the FIR would be sensitive to such spectral features, as well.

\begin{figure*}
        \begin{center}
\plottwo{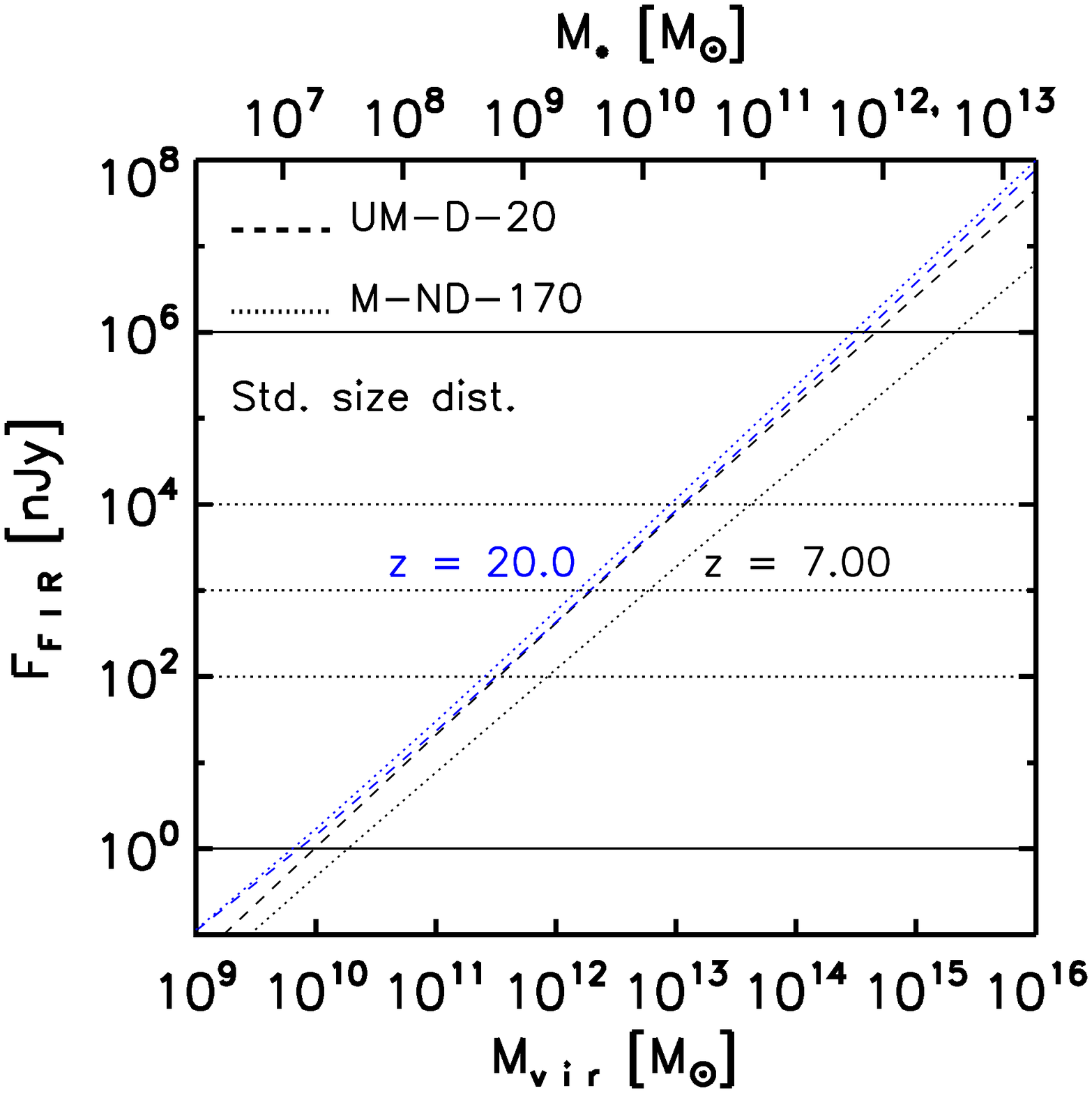}{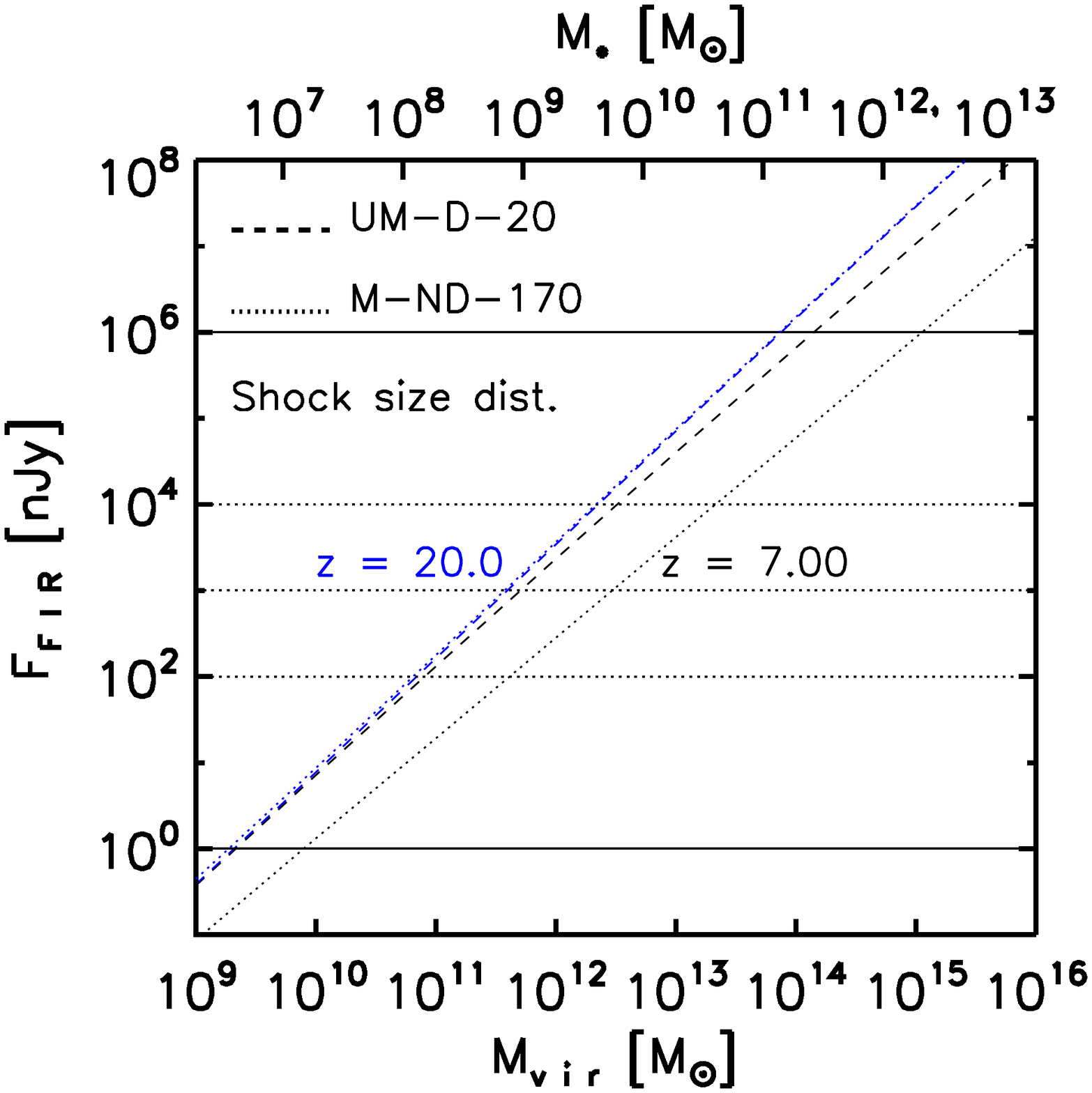}
        \end{center}
    \caption{
	    FIR flux as a function of $M_{\rm vir}$ (bottom horizontal axis) and $M_{\rm *}$ 
	    (top horizontal axis) for the same dust models
	    shown in Fig. \ref{fig:Lnu_fnu_vs_nu} in the case of a standard (left-hand panel) 
	    and shock (right-hand panel) size distributions. Results for $z=7$ (black lines) 
	    and $z=20$ (blue lines) are
	    compared.  Horizontal dashed lines denote select sensitivity values 
	    ($0.1$, $1$ and $10 \ {\mu}{\rm Jy}$) 
	    that will be considered for analysing the potential detection of first galaxies. Horizontal solid
	    lines enclose the whole range of sensitivity values considered in this work.
    }

    \label{fig:Ffir_vs_M}
\end{figure*}

In order to more fully explore the effects of the strong negative K-correction affecting primeval galaxies,
Fig. \ref{fig:Ffir_vs_z} compares the FIR flux predicted by our model (thick lines) with
the FIR flux obtained assuming that model galaxies with similar
masses exhibit the same ($z=7$) rest-frame specific luminosity at all redshifts (thin lines).
Results for galaxies of different masses ($10^{10}$ and $10^{11} \ {\rm M}_{\sun}$, corresponding 
to $M_{\rm *} \approx 1.8\times 10^{7}$ and $1.8\times10^{8} \ {\rm M}_{\sun}$, respectively)
are shown and correspond to the same dust models analysed in Fig. \ref{fig:Lnu_fnu_vs_nu}.
In the case of the UM-D-20 model, which leads to weaker spectral features (Fig. \ref{fig:Lnu_fnu_vs_nu}),
the negative K-correction generates an almost constant FIR flux at $z=7-20$, while, for the
M-ND-170 model, associated to stronger spectral features, the flux remains almost constant at $z=12-20$ and decreases by a factor of 2 towards $z=7$ (thick lines).  
In the latter case, the higher FIR fluxes at $z=12-20$ is explained by the main spectral features
that enter the FIR filter ($\Delta \lambda = 250-750 \ {\mu}{\rm m}$) at $z\ga 12$ (see Fig. \ref{fig:Lnu_fnu_vs_nu}).
If K-correction effects were not present (thin lines), the FIR flux associated to the UM-D-20 (M-ND-170) model
would decrease by a factor of $\sim 10$ ($\sim 2$) between $z=7$ and $20$.
At $z=20$, it is clear that the strong negative K-correction leads to an increase of the observed flux
by $\sim 1$ order of magnitude for both considered models.
Similar trends are obtained for the whole range of 
masses associated to first galaxies ($M_{\rm vir} > 10^7 \ {\rm M}_{\sun}$).

In Fig.~\ref{fig:Ffir_vs_M}, we analyse the FIR flux as a function of $M_{\rm vir}$ {and $M_{*}$} for the same dust 
chemical models used in Fig. \ref{fig:Lnu_fnu_vs_nu}.  Left- and right- hand panels show results for 
a standard and shock size distributions of dust grains, respectively.  Predictions for 
$z=7$ (black lines)
and $z=20$ (blue lines) are compared.  Intermediate trends are obtained at intermediate redshifts
($7<z<20$). 
As expected from Fig.~\ref{fig:Ffir_vs_z}, the $F_{\rm FIR}-M_{\rm vir}$ relation
associated to the UM-D-20 dust model seems not to evolve with $z$; i.e., systems of a given
mass present similar observed fluxes, regardless of their redshift.  On the other hand,
the $F_{\rm FIR}-M_{\rm vir}$ relation corresponding to the M-ND-170 model moves towards lower fluxes
at lower $z$ (see also Fig.~\ref{fig:Ffir_vs_z}); thus, contrary to expectations, 
systems with similar masses are brighter if they are located at higher $z$.
Thus, a strong negative K-correction is also evident here.
Horizontal dashed lines in Fig.~\ref{fig:Ffir_vs_M} denote select characteristic sensitivity values
($S = 0.1$, $1$ and $10 \ {\mu}{\rm Jy}$), covering the scope of future FIR surveys.
We see that masses of 
$M_{\rm vir}\ga 10^{11}-10^{12} \ {\rm M}_{\sun}$ and $M_{\rm *}\ga 1.8\times10^{8}- 1.8\times10^{9} \ {\rm M}_{\sun}$ ($M_{\rm vir}\ga 10^{13}-10^{14} \ {\rm M}_{\sun}$ and 
$M_{\rm *}\ga 1.8\times 10^{10}- 1.8 \times 10^{11} \ {\rm M}_{\sun}$) 
are required to reach sensitivity limits of $S \sim 0.1 \ {\mu}{\rm Jy}$ ($S\sim 10.0 \ {\mu}{\rm Jy}$),
with the exact value depending on dust properties.  In the next section, we will discuss the minimum mass 
required for detecting primeval sources in greater detail.  

\begin{figure*}
	\begin{center}
\plottwo{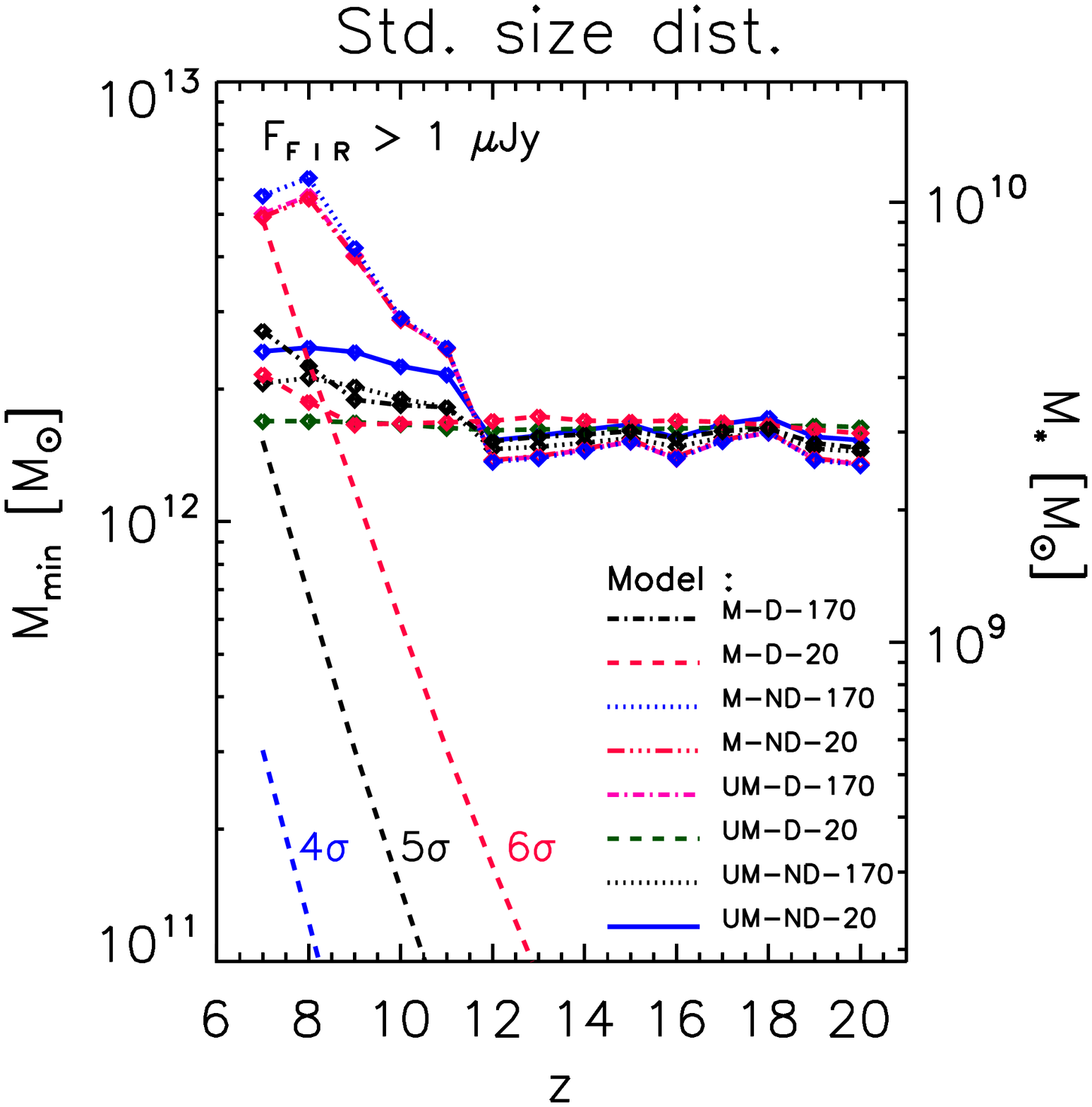}{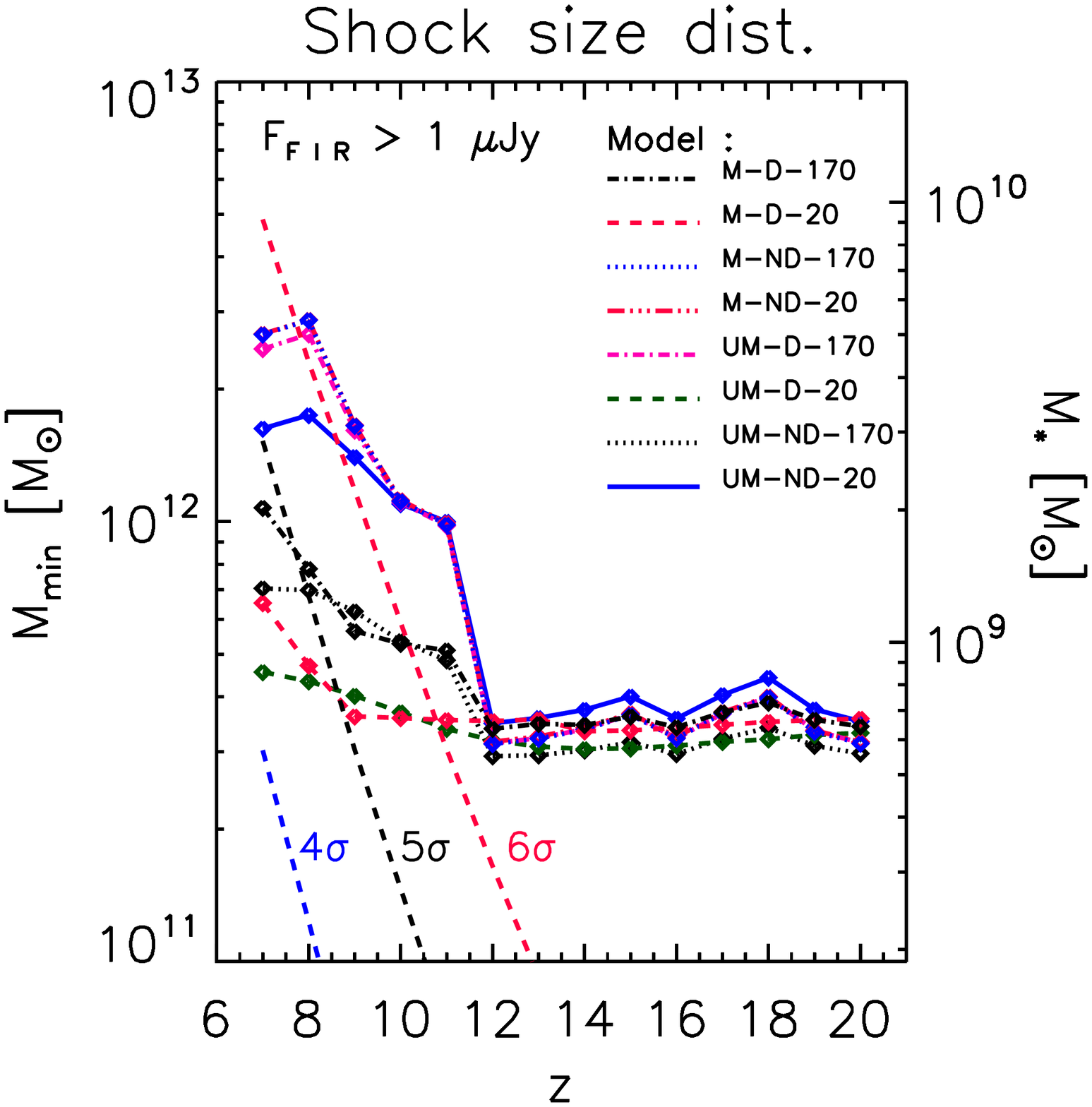}
	\end{center}
    \caption{
	    Lowest virial mass (left vertical axis) required for detection as a function of $z$, assuming a sensitivity 
	    of $1 {\mu}{\rm Jy}$.
	 Right vertical axes show the associated stellar mass values
	adopting our conservative star formation efficiency $\eta = 0.01$.
            Predictions for different dust
models are shown: standard (left) and shock (right) dust size distributions,
as indicated in the figure. 
For comparison, different $\nu$-$\sigma$ peaks corresponding to the adopted
	cosmological model are plotted.  
	}
    \label{fig:Mmin_vs_z}
\end{figure*}

\section{Minimum mass limit}
\label{sec:min_mass}

In Section~\ref{sec:k_correction}, we found that primeval FIR/sub-mm sources experience a significant negative K-correction. In this section, we determine the minimum galaxy mass required for detection with a generic FIR telescope, as a function of $z$ and for different dust models.

In Fig.~\ref{fig:Mmin_vs_z}, we predict the virial mass limit, $M_{\rm min}$, 
for detections as a function of $z$ (see Sec.~\ref{sec:methodology}). 
         As a reference, right vertical axes show the associated stellar mass values
        adopting our conservative star formation efficiency $\eta = 0.01$.
Predictions of different dust
models for the standard (left) and shock (right) dust size distributions are shown,
as indicated in the figure. An instrument sensitivity of $1 \, \mu {\rm Jy}$ at
${\lambda}_{\rm obs} = 250-750 \, {\mu}{\rm m}$ has been assumed, which is a plausible
value for a potential ultra-deep blind survey in the next decades.
For the sake of comparison, we show curves depicting halo masses for different overdensities ($\nu$-$\sigma$ peaks) in the underlying Gaussian field of cosmological density fluctuations.

For clarity, Fig.~\ref{fig:Mmin_vs_z} does not show the $M_{\rm min}-z$ relations for different sensitivity values, but such scalings can be easily inferred by combining data from Fig.~\ref{fig:Ffir_vs_M} and \ref{fig:Mmin_vs_z}. 
We note that, according to Fig.~\ref{fig:Mmin_vs_z}, dust models UM-D-20 and M-ND-170
lead to the lowest and highest $M_{\rm min}$ values for a given $z$ and size distribution, with
all other models exhibiting an intermediate behaviour. 
Furthermore, we find that, in general, different models exhibit similar $M_{\rm min}-z$ curves for different
sensitivity values, but with an increase of the $M_{\rm min}$ normalization for higher sensitivities.
Thus, for models UM-D-20 and M-ND-170, the $M_{\rm min}$ normalization can be derived 
from Fig.~\ref{fig:Ffir_vs_M} by determining where
the corresponding $F_{\rm FIR}-M_{\rm vir}$ curves (at $z=7$ and 20) cross the horizontal line depicting 
a given sensitivity $S$.  
As models UM-D-20 and M-ND-170 represent upper and lower limits for $M_{\rm min}$, 
the $M_{\rm min}$ normalization (associated to sensitivity $S$),
for the remaining chemical models,
would lie approximately between these two extreme cases.

Comparing left- and right-hand panels in Fig.~\ref{fig:Mmin_vs_z}, we note that $M_{\rm min}$ 
is lower for the shock size distribution of dust grains, which is explained by the higher $F_{\rm FIR}$ fluxes
obtained in this case (Fig.~\ref{fig:Ffir_vs_M}).  As discussed in \citet{derossi2017},
the shock size distribution leads to higher $T_{\rm d}$ and dust emissivities, which is consistent 
with the aforementioned behaviour.
In the case of the standard size distribution, $M_{\rm min} \approx 1-2\times10^{12} \ {\rm M}_{\sun}$
at $z \ga 12$, while for the shock size distribution, $M_{\rm min} \approx 3-4\times10^{11} \ {\rm M}_{\sun}$
at similar $z$. The extreme negative K-correction, as noted previously, leads to a
lowest value of $M_{\rm min}$ at $z\ga12$. 
In addition, we find another unexpected behaviour, seen at $z\la12$: $M_{\rm min}$ increases towards
lower $z$, depending on the dust model.
In the case of the standard size distribution, $M_{\rm min}$ reaches a maximum of 
$\approx 6\times10^{12} \ {\rm M}_{\sun}$ at $z\la8$, when applying the M-ND-170 dust chemistry.
For the same dust chemistry, but implementing the shock size distribution,  $M_{\rm min}$ reaches its maximum of
$\approx 3\times10^{12} \ {\rm M}_{\sun}$ at $z\la8$.
However, other dust chemistries, such as the UM-D-20 model, only show modest variations
of $M_{\rm min}$ at $z\la12$.

\begin{figure*}
        \begin{center}
\plottwo{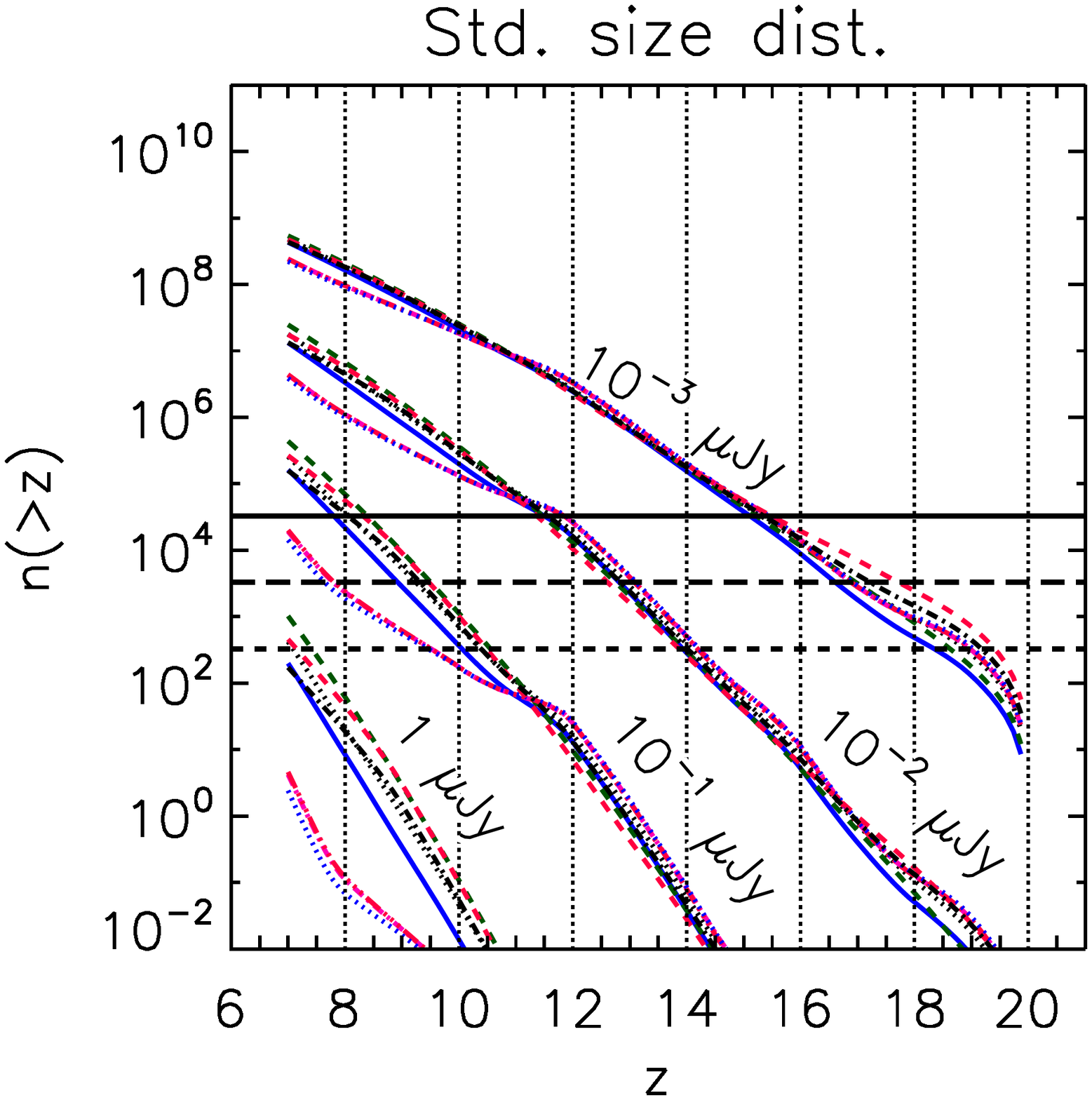}{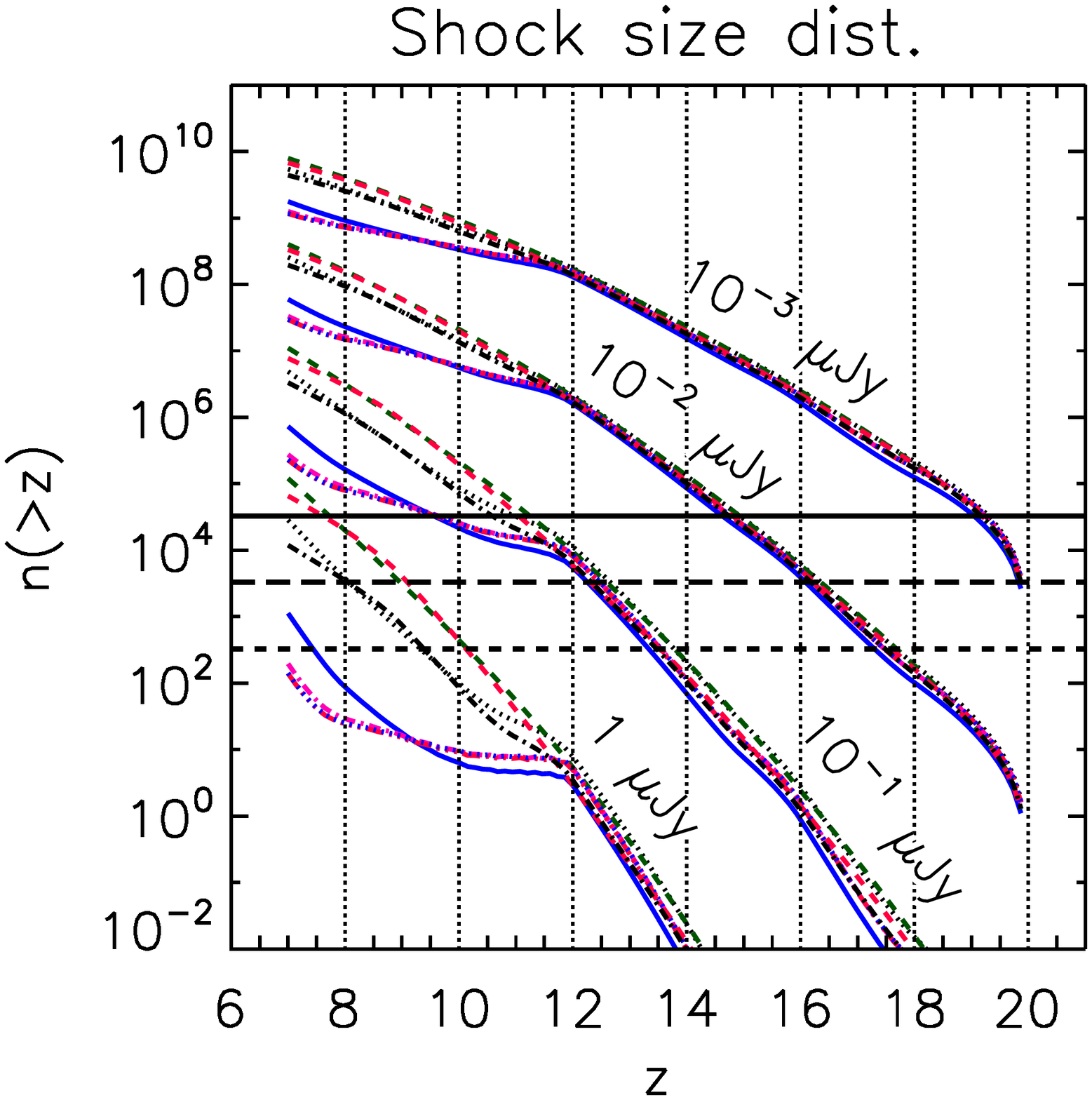}
        \end{center}
    \caption{
	    Number of sources per steradian above redshift $z$ detected during
        potential surveys. Results for different sensitivity values are explored, as indicated in the figure.
Results for the standard (left-hand panel)
and shock (right-hand panel) grain size distributions are shown. Different dust models (see Fig.
\ref{fig:Mmin_vs_z} for line style convention) are considered. 
	The horizontal lines depict the regions corresponding to the redshift horizon
	($n(>z_{\rm lim}) = 1 / {\Delta}{\Omega}$)  considering survey areas of 
	${\Delta}{\Omega} = 0.1$ (solid line), $1.0$ (long-dashed line) and $10 \ {\rm deg}^2$ (dashed line), respectively. 
        }
    \label{fig:n_vs_z}
\end{figure*}

The peculiar trends obtained at $z\approx7-12$ can be explained again (see also Section~\ref{sec:k_correction}) 
by analysing the predicted spectra shown in Fig. \ref{fig:Lnu_fnu_vs_nu}.  There, we compare the energy distributions
associated with the UM-D-20 and M-ND-170 models. The former exhibit the lowest $M_{\rm min}$
at $z=7$, and the latter the highest $M_{\rm min}$ at the same $z$ (Fig. \ref{fig:Mmin_vs_z}).
By comparing the spectra corresponding to these two extreme cases, it is clear that differences
between the average fluxes ($F_{\rm FIR}$) are driven by the different redshifted spectral
features that enter the wavelength range corresponding to our fiducial FIR filter
($\Delta \lambda = 250 - 750 \ {\mu}{\rm m}$, vertical dashed black lines in Fig. \ref{fig:Lnu_fnu_vs_nu}) 
at a given $z$.
At $z>12$, both dust models exhibit similar values of $M_{\rm min}$
because, at those $z$, all the main spectral features have
entered the adopted FIR band.  
At $z<12$, the average flux evolves, as different spectral features
enter the FIR band.  The model UM-D-20, which exhibits weaker features, does
not evolve significantly over the whole $z$ range.  On the other hand, the model M-ND-170, which exhibits
stronger features, shows more significant evolution at $7 < z < 12$.
The dust spectral features are associated with the underlying dust opacities, and therefore depend on dust chemical composition.
According to these findings, the chemical composition of dust in primordial galaxies at $z=7-12$
might affect their detection at FIR/sub-mm wavelengths. Future telescopes
operating in this wavelength range and with low enough sensitivities could, thus, constrain 
the nature of dust in the early universe.

Although the strong negative K-correction, found for FIR/sub-mm sources at early times,
enhances their chances of being detected, halos with 
$M_{\rm vir} > M_{\rm min} > 10^{11} \ {\rm M}_{\sun}$ are extremely rare at high $z$.
To illustrate this fact, the $M_{\rm vir}-z$ 
curves for different $\nu$-$\sigma$ overdensities are overplotted in Fig. \ref{fig:Mmin_vs_z}.  
We see that systems with $M_{\rm vir} > M_{\rm min}$ 
are far from typical at those redshifts, representing highly biased overdensities.
Fig.~\ref{fig:Mmin_vs_z} shows results for a sensitivity of 1 ${\mu}{\rm Jy}$. Even if we consider extreme sensitivities of 1~nJy, a minimum mass of $\ga 10^9 \ M_{\sun}$
(see Fig.~\ref{fig:Ffir_vs_M}) would be required for detection, above a $2-{\sigma}$ peak at these high redshifts.
As our model sources represent the first galaxies hosting
primordial dust grains, their dust masses are very low and high masses
are required to achieve detectable fluxes at FIR/sub-mm wavelengths \citep[e.g.,][]{jaacks2018}.
Thus, although possible, detecting these systems in the dust continuum would be difficult with blind surveys.
However, we acknowledge significant uncertainties regarding the nature of primeval
galaxies and their dust content. In Section~\ref{sec:parameter_variations},
we show that an increase of our adopted values of $D/M$, $Z_{\rm g}$ or $\eta$ would lead
to lower values of $M_{\rm min}$, enhancing the chances of detecting the dust emission
from first galaxies.
As discussed below, gravitational lensing  could also be a crucial tool for detecting the most
elusive faint sources, and such analysis will be the subject of a forthcoming study.

Next, we analyse the probability of individual detections during potential blind surveys
considering different instrument sensitivities and survey areas.

\section{The redshift horizon}
\label{sec:z_horizon}

Following the procedures described in Section~\ref{sec:density_model}, in this section, we derive
the projected number of sources (Equ. \ref{eq:N_z}) as a function of $z$ 
during potential blind surveys, assuming observations with different instrument sensitivities and 
considering different survey areas.
Note that we only consider redshifts ranging from 7 to 20, 
as these are the plausible times when the first galaxies could have formed.

Fig. \ref{fig:n_vs_z} shows the number of sources per steradian above redshift $z$ ($n(>z)$)
which could be detected during potential surveys. 
Results for the standard (left-hand panel)
and shock (right-hand panel) grain size distributions are shown and different dust chemical 
models (see Fig. \ref{fig:Mmin_vs_z} for line style convention) are considered.
The horizontal lines depict the regions corresponding to the redshift horizon
($n(>z_{\rm lim}) = 1 / {\Delta}{\Omega}$) considering survey areas of
${\Delta}{\Omega} = 0.1$ (solid line), $1.0$ (long-dashed line) and $10 \ {\rm deg}^2$ (dashed line), respectively.
Results for different sensitivity values are explored:
$S = 1, \ 10^{-1}, \ 10^{-2}$ and $10^{-3} \ {\mu}{\rm Jy}$.  
Curves associated with sensitivities greater than $\sim 1 \ {\mu}{\rm Jy}$ do not reach the horizon
for the adopted survey areas; hence, for the sake of clarity, these curves were not plotted.

\begin{figure*}
        \begin{center}
\plottwo{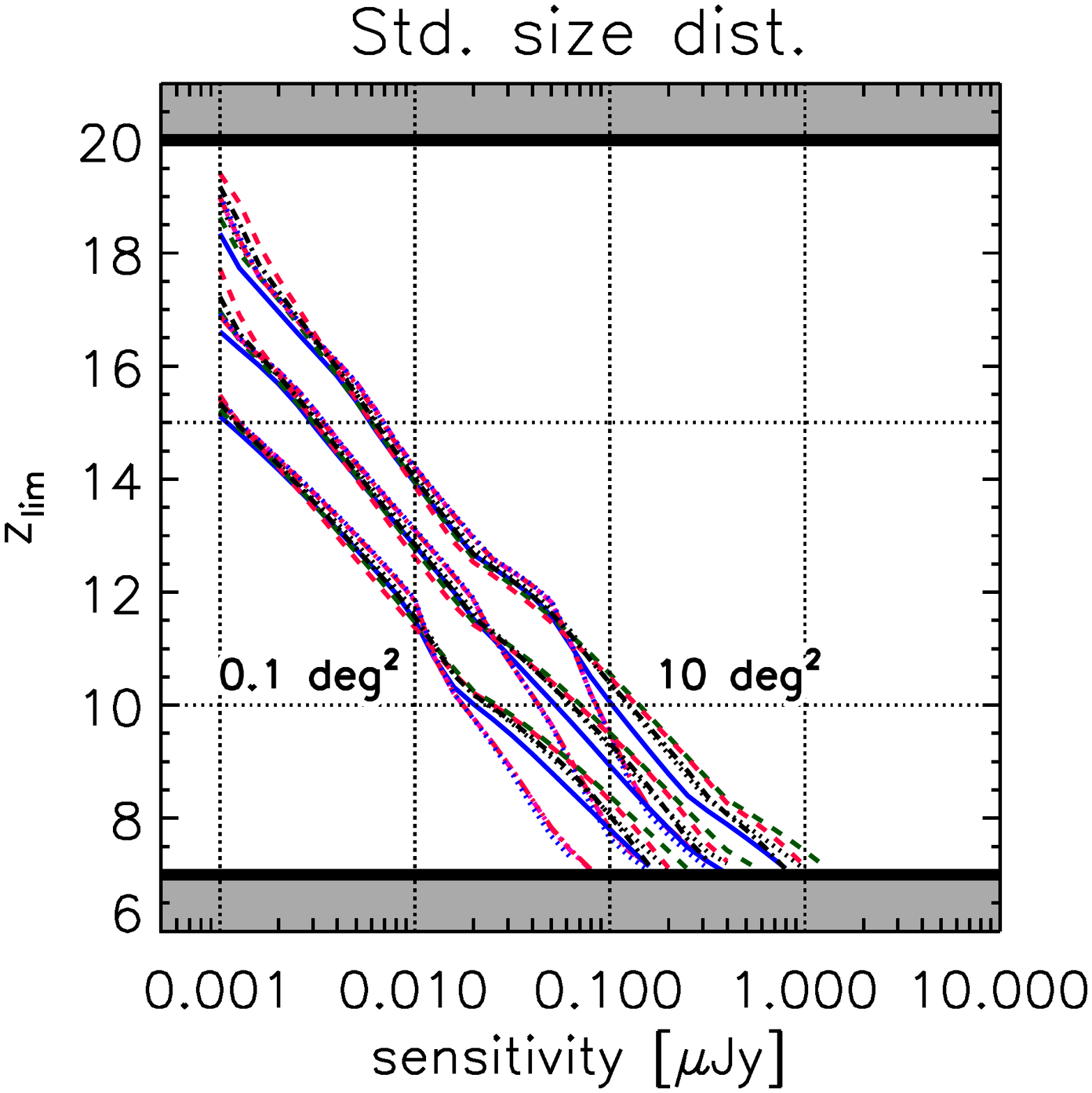}{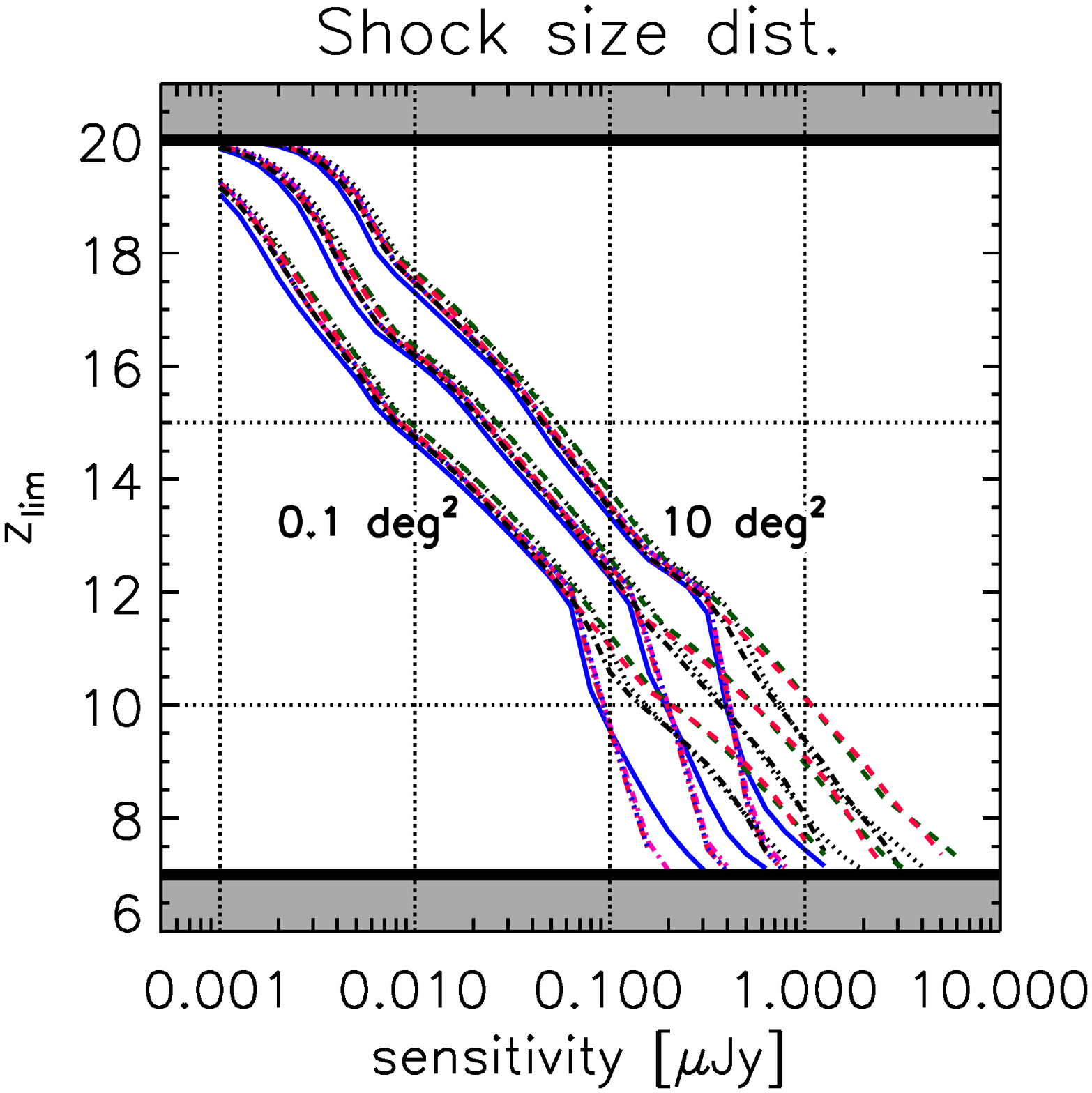}
        \end{center}
    \caption{
	    Redshift horizon as a function of instrument sensitivity
	    for three different survey areas: 0.1, 1.0 and 10.0 deg$^2$, as indicated in the figure.
	    Results for the standard (left-hand panel)
and shock (right-hand panel) grain size distributions are shown. Different dust models (see Fig.
\ref{fig:Mmin_vs_z} for line style convention) are considered. Our analysis is restricted to 
	$z=7-20$, where first galaxies are expected to form.
        }
    \label{fig:redshift_horizon}
\end{figure*}

As expected from Fig. \ref{fig:Mmin_vs_z}, Fig. \ref{fig:n_vs_z} shows that
higher source densities are obtained 
for the shock size distribution, reaching differences of more than one order of magnitude. 
Regarding the dust chemical composition, it also affects source densities, with a
larger impact when the shock size distribution is adopted.
In particular, we see that, for a given size distribution and sensitivity $S$, 
different dust chemistries lead to similar $n(>z)-z$ curves at $z>12$.  On the
other hand, at $z<12$, $n(>z)$ exhibits a more
significant dependence on the dust chemical composition.  The latter
behavior is a consequence of the dependence of $M_{\rm min}$ on the dust
chemistry at $z<12$ (see Sec.~\ref{sec:min_mass}), which is caused by the different
spectral features that enter the adopted FIR filter at $z<12$ for different models.
Thus, our findings suggest that the number of detected sources depends on the
properties of primordial dust and, hence, measurements of source densities 
might provide important constraints on the nature of dust chemical composition and 
grain size distributions at very early cosmic times.

According to Fig.~\ref{fig:n_vs_z}, the probability
of detecting one individual source exhibiting {\em typical} properties
expected for first galaxies depends mainly on the instrument sensitivity
and, secondly, on the survey area. 
In the case of the shock size distribution, the redshift horizon
for $S\sim 1 \ {\mu}{\rm Jy}$ is reached at $z\sim 7-10$, with
the exact value depending on ${\Delta}{\Omega}$ and dust chemical composition.
In the case of the standard size distribution, the redshift horizon
for $S\sim 1 \ {\mu}{\rm Jy}$ only reaches $z\sim7$ for 
${\Delta}{\Omega}=10\ {\rm deg}^2$ and for some particular dust chemistries.
For instrument sensitivities $S \la 0.1 \ {\mu}{\rm Jy}$ and 
${\Delta}{\Omega} \ga 0.1 \ {\rm deg}^2$, all dust models
predict a redshift horizon above $z\approx7$.

Fig.~\ref{fig:redshift_horizon} summarizes our findings regarding
the redshift horizon ($z_{\rm lim}$) as a function of sensitivity 
($S$) for different survey areas and dust models.  
For the standard size distribution, the $z_{\rm lim}-{\log}_{10}(S)$ relation
at $z_{\rm lim}>12$, approximately follows a relation of the form 
$z_{\rm lim} = -4 \ {\log}_{10}(S/{\rm nJy}) + z_{\rm lim}(S=1\ {\rm nJy})$, where
$z_{\rm lim}(S=1 \ {\rm nJy})\approx 15,\ 17$ and 18, for $\Delta \Omega = 0.1, 1$
and $10 \ {\rm deg}^2$, respectively. A similar relation, with roughly
the same slope, is obtained for the shock size distribution at $z_{\rm lim}>12$, but with
$z_{\rm lim}(S=1 \ {\rm nJy})\approx 19,\ 20$ and 21.5, for $\Delta \Omega = 0.1, 1$
and $10 \ {\rm deg}^2$, respectively.
At $z<12$, the $z_{\rm lim}-{\log}_{10}(S)$ curves depart from linearity for some dust models
as their associated spectral features enter our adopted FIR filter (see Sec.~\ref{sec:min_mass}).
Only models with weak spectral features (e.g. UM-D-20) predict a linear behavior along
the whole redshift range associated with first galaxy formation.

Considering our whole set of plausible dust models,  
a sensitivity
$S\la0.1 \ {\mu}{\rm Jy}$ and a survey area $\Delta \Omega \ga 0.1 \ {\rm deg}^2$ are required
in order to 
assure the detection of the dust continuum emission
of at least one individual first galaxy during a FIR survey at $z>7$.
However, depending on the exact properties of primordial dust, a redshift horizon
$z_{\rm lim}\ga 7$ could be also achieved 
for sensitivities $S\la10 \ {\mu}{\rm Jy}$ ($S\la1 \ {\mu}{\rm Jy}$) and
$\Delta \Omega \ga 10 \ {\rm deg}^2$ ($\Delta \Omega \ga 0.1 \ {\rm deg}^2$).
Finally, only in the case of a shock size distribution of dust grains, the maximum redshift horizon 
$z_{\rm lim}\approx20$ could
be reached for an extreme sensitivity $S\sim 1$ nJy and $\Delta \Omega = 0.1 - 10 \ {\rm deg}^2$.

Although our findings suggest that
the probability of detecting
{\em typical} first galaxies in blind surveys will be very challenging,
given the extreme sensitivities required,
the nature of primordial galaxies is still uncertain and
we will show that observational prospects increase significantly with the metallicity
and dust-to-metal ratio of the considered model sources (Section~\ref{sec:parameter_variations}).

\begin{figure*}
        \begin{center}
\plottwo{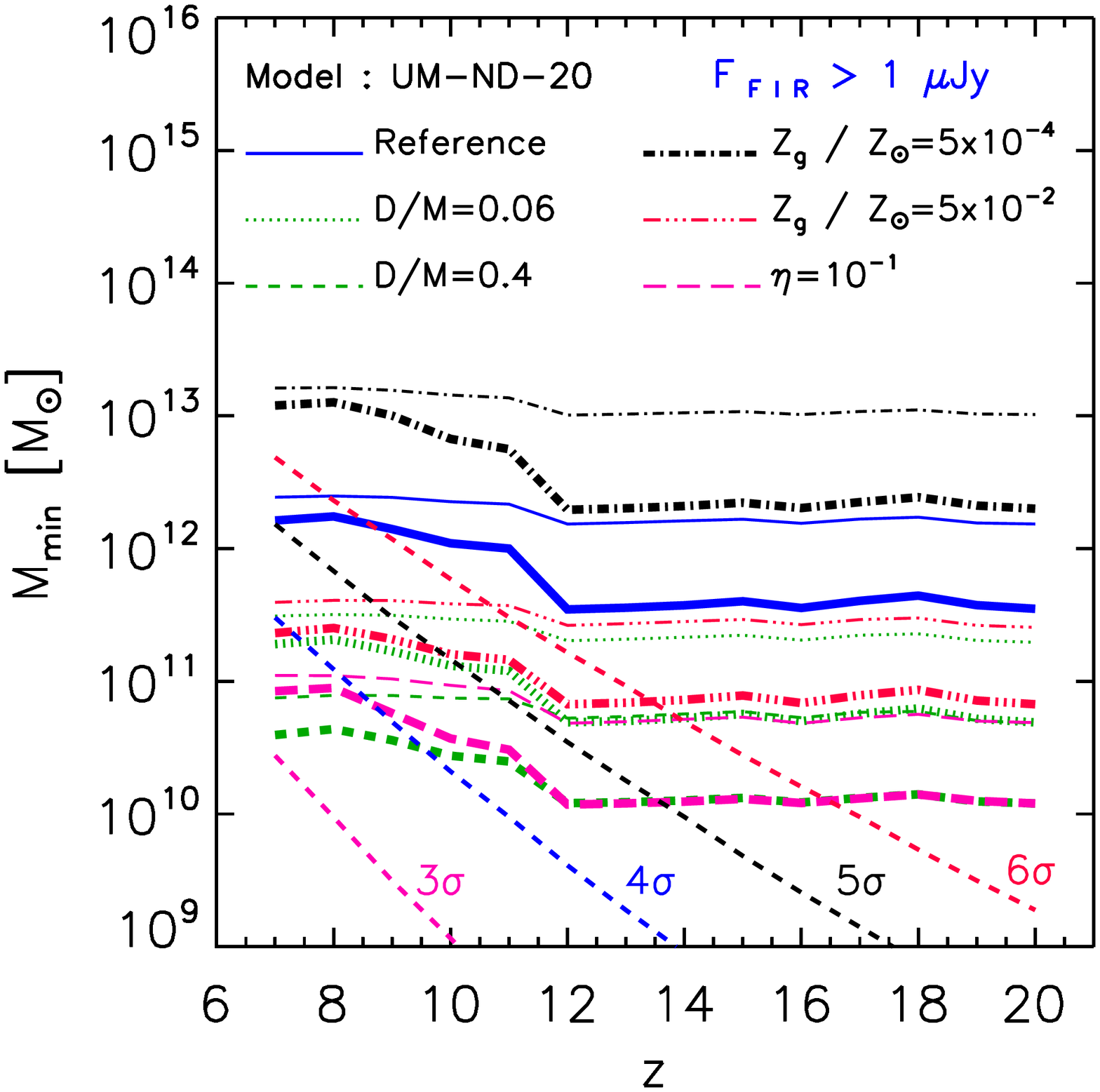}{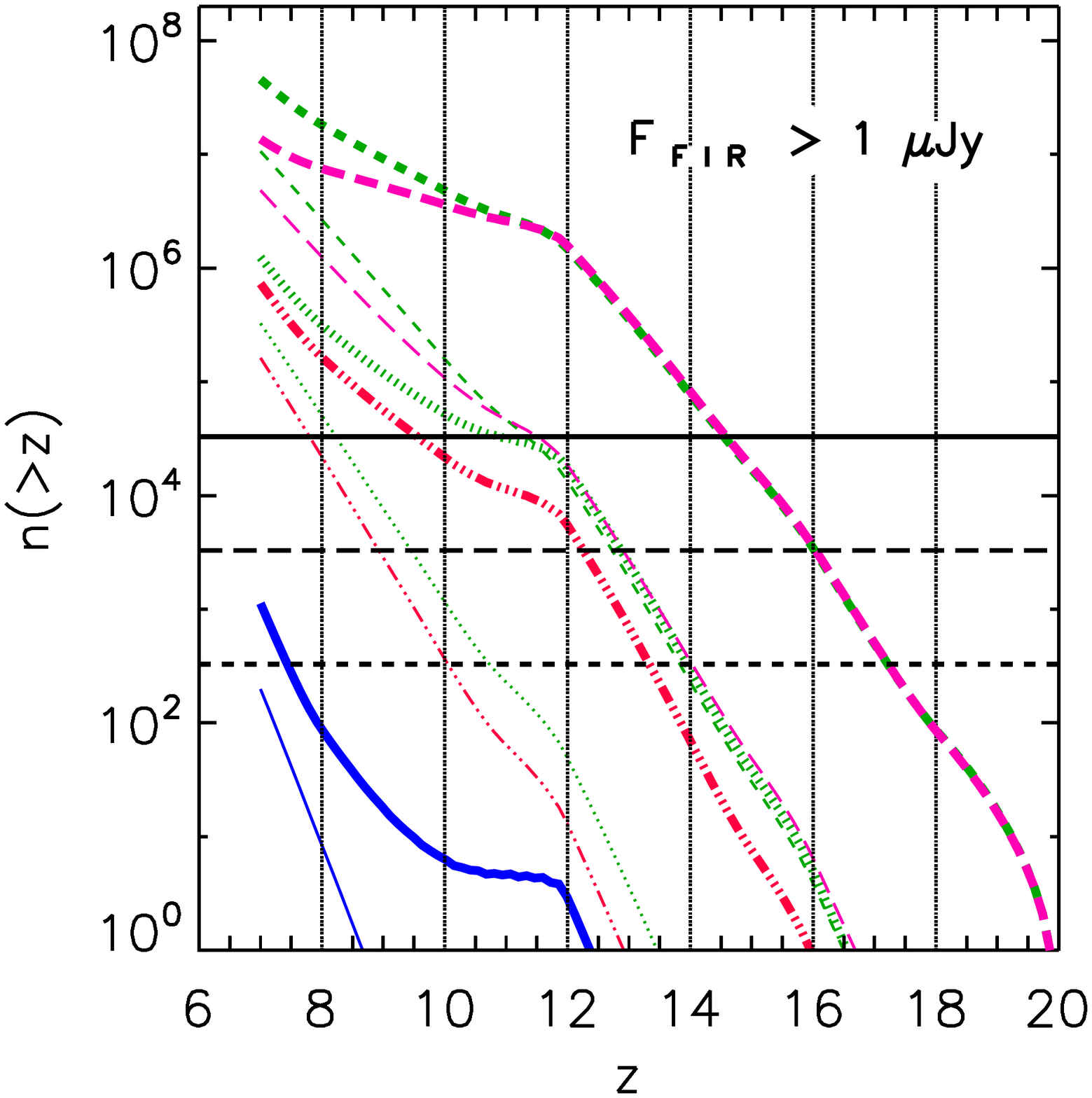}
        \end{center}
    \caption{
Impact of varying model parameters on the $M_{\rm min}-z$ (left-hand panel)
and $n-z$ (right-hand panel)
	relations, adopting a sensitivity $S = 1 \ {\mu}{\rm Jy}$ (compare with Fig. \ref{fig:Mmin_vs_z} and \ref{fig:n_vs_z}, respectively).
We only show results for the UM-ND-20 model, to emphasize the key features.
Thin and thick lines are associated with the standard and shock size distributions, respectively.
The reference model (solid blue line) corresponds to our standard parameters \citep{derossi2017}.
The effects of variations of model parameters with respect to the standard model are shown with different line styles.
Note that, in the right-hand panel, the curves corresponding to the case
	$Z_{\rm g} = 5\times10^{-4}\ {\rm Z}_{\sun}$ (dot-dashed lines) are not shown as they lie below the plotted region. For $\eta =  0.1$, $M_* / M_{\rm min} \approx 2 \times 10^{-2}$, while,
	for the rest of the models, $M_* / M_{\rm min} \approx 2 \times 10^{-3}$.
        }
    \label{fig:variations_Mmin_n}
\end{figure*}

\section{Parameter sensitivity}
\label{sec:parameter_variations}

The nature of primordial dust and its host galaxies is 
still a matter of debate.
In order to estimate the uncertainties of our model predictions, 
we evaluate the effects of varying the more critical parameters.
In the following, for the sake of clarity, 
only results for the UM-ND-20 model are shown as this model exhibits an
intermediate behavior (see Fig.\,\ref{fig:Mmin_vs_z} and \ref{fig:n_vs_z}).
All other models predict similar trends but with slight variations
in the normalization of $M_{\rm min}$ and $n(>z)$ at $z<12$.
In Fig.~\ref{fig:variations_Mmin_n}, we show the impact of variations of different model 
parameters on the $M_{\rm min}-z$ (left-hand panel) and $n-z$ (right-hand panel) relations.
For the sake of comparison with Fig. \ref{fig:Mmin_vs_z} and \ref{fig:n_vs_z}, 
we adopted an instrument sensitivity $S=1\ {\mu}{\rm Jy}$.  As discussed before,
if varying the sensitivity, similar curves are obtained but with a change in the
absolute normalization of the relations (Figs.~\ref{fig:Ffir_vs_M}, \ref{fig:Mmin_vs_z}
and \ref{fig:n_vs_z}).

For evaluating the effects of the dust-to-metal ratio, we consider the cases:
$D/M=$ 0.005 (reference), 0.06 and 0.4 \citep{cen2014}.  We can see that $M_{\rm min}$
decreases almost proportionally to the increase of $D/M$.  
This is consistent with the results in \citet{derossi2017}, who reported that
the dust radiation intensity increases nearly linearly with the dust fraction.
In particular, for the extreme value of $D/M=0.4$ and a shock size distribution,
$M_{\rm min} = 1-10 \times 10^{10} \ {\rm M}_{\sun}$.
We also tested the effects of changing the gas metallicity to 
$Z_{\rm g}= 5\times10^{-4} ,~5\times10^{-3} \ {\rm (reference) \ and \ 5\times10^{-2}}  \ {\rm Z}_{\sun}$.
As noted by \citet{derossi2017}, $D/M$ and $Z_{\rm g}$ are degenerate parameters,
with $M_{\rm min}$ increasing almost proportionally to the decrease of $Z_{\rm g}$.
For the extreme lowest value of $Z_{\rm g} \approx 5\times10^{-4} \ {\rm Z}_{\sun}$ and a standard
distribution for dust grains, $M_{\rm min}$ reaches 
the highest value of $\approx 10^{13} \ {\rm M}_{\sun}$.
For $Z_{\rm g} \approx 5\times10^{-2} \ {\rm Z}_{\sun}$, we obtain similar changes to those
derived from using $D/M=0.06$.  To explore the impact of the adopted star formation
efficiency, we increased ${\eta}$ to the extreme value of 0.1.
We note that, for $\eta =  0.1$,  the stellar-to-total mass ratio is 
$M_* / M_{\rm min} \approx 2 \times 10^{-2}$, while, as previously discussed,
        for the rest of the models explored in this
	work, $M_* / M_{\rm min} \approx 2 \times 10^{-3}$.
As a higher ${\eta}$ would drive an enhanced ISM metallicity, we also increased
$Z_{\rm g}$ to $0.05 \ {\rm Z}_{\sun}$, consistent with the closed-box model approximation:
$Z = - y \ln ( 1 - {\eta})$, where $y$ is the stellar yield. Although this
approximation might not describe the real behavior of these sources,
it provides an upper limit for the predicted dust emission.
We see that these changes lead to a lower $M_{\rm min}$, obtaining similar results
to those derived when increasing $D/M$ to the extreme value of 0.4.
In summary, variations of model parameters to extremely low and high values lead to 
significant changes in $M_{\rm min}$, resulting in variations of more than two orders of magnitude.

With respect to the number density of sources, $n(>z)$, a similar sensitivity is seen
when varying model parameters to extreme values.
For $Z_{\rm g}= 5\times10^{-4} \ {\rm Z}_{\sun}$, 
the probability of detecting one primeval source, assuming $S\sim 1 \ {\mu}{\rm Jy}$
and $\Delta \Omega \la 10 \ {\rm deg}^2$,
is below unity ($n(>z) < 1/\Delta \Omega$) at all redshifts $z \ge 7$.
In the case of our default metallicity value ($Z_{\rm g}= 5\times10^{-3} \ {\rm Z}_{\sun}$)
and adopting $\Delta \Omega \la 10 \ {\rm deg}^2$,
this probability increases to 1 at $z\approx7$.
On the other hand, for $D/M\ga0.06$, $Z_{\rm g} \ga 5\times10^{-2} \ {\rm Z}_{\sun}$ or ${\eta}=0.1$,
and assuming $S\sim 1 \ {\mu}{\rm Jy}$,
the redshift horizon, defined as the redshift where $n(>z)=1/\Delta \Omega$, 
extends towards $z\ga7$ for our whole range of considered survey areas ($\Delta \Omega = 0.1-10 \ {\rm deg}^2$). 
In the case of a standard size distribution and for $D/M\sim0.06$ or 
$Z_{\rm g} \sim 5\times10^{-2} \ {\rm Z}_{\sun}$, the redshift horizon is reached
at $z\approx8,\ 9$ and 10 for $\Delta \Omega \approx 0.1, \ 1$ and $10 \ {\rm deg}^2$, respectively.  
In the case of a shock (standard) size distribution 
and an extreme value of $D/M\sim0.4$ or ${\eta}=0.1$, 
the redshift horizon extends 
towards $z\sim15, \ 16$ and 17 ($z\sim 12, \ 13$ and 14) for 
$\Delta \Omega \approx 0.1, \ 1$ and $10 \ {\rm deg}^2$, respectively.
All other considered cases exhibit intermediate behaviors.
Note that, if we had considered other dust chemical compositions 
(e.g. UM-D-20, M-D-20, UM-ND-170, M-D-170, instead of UM-ND-20) which lead to lower $M_{\rm min}$ 
(see Fig.~\ref{fig:Mmin_vs_z}), the redshift horizon would have moved towards
even higher $z$.

We conclude that the redshift horizon at FIR wavelengths strongly depends 
on the properties of primordial dust and its host galaxy population.
For $S=1 \ {\mu}{\rm m}$ and assuming a UM-ND-20 dust chemistry, metallicities 
$Z_{\rm g} \ga 5\times10^{-2} \ {\rm Z}_{\sun}$ 
or dust fractions $D/M\ga0.06$ are required
to reach the redshift horizon at $z > 7$ for $\Delta \Omega=0.1-10 \ {\rm deg}^2$. 
However, these limits for $Z_{\rm g}$ and $D/M$ could move 
towards lower values for different dust chemical compositions.  
For galaxies below such metallicity and dust fraction limits,
gravitational lensing would be required to detect these fainter systems if sensitivity limits cannot
be improved.  We plan to explore the possibility of detecting gravitational lensing
systems in a future study.

\begin{figure}
\plotone{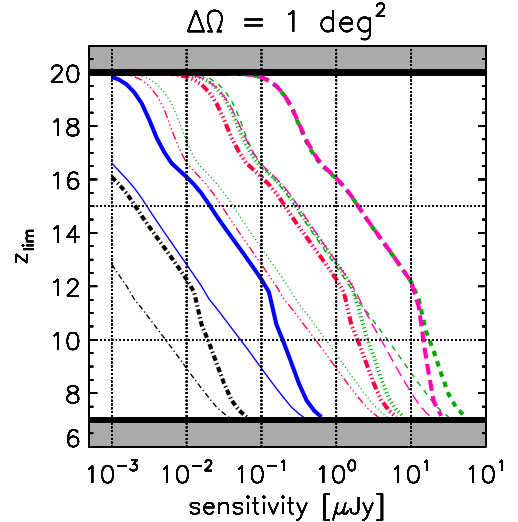}
        \caption{
		Impact of variations of different model parameters on the redshift horizon
	(compare with Fig. \ref{fig:redshift_horizon}) for a survey area of $1 \ {\rm deg}^2$.
We only show results for the UM-ND-20 model, to emphasize the key features.
Thin and thick lines are associated with the standard and shock size distributions, respectively.
The reference model (solid blue line) corresponds to our standard parameters \citep{derossi2017}.
	The effects of variations of model parameters with respect to the standard model are shown with different line styles (see Fig. \ref{fig:variations_Mmin_n}, for line style convention).
}
    \label{fig:variations_redshift_horizon}
\end{figure}

Fig.~\ref{fig:variations_redshift_horizon} summarizes our results regarding the
redshift horizon as a function of sensitivity for a survey area of $1 \ {\rm deg}^2$,
considering the different galaxy parameters
explored in Fig.~\ref{fig:variations_Mmin_n}.
It is evident that an increase by one order of magnitude of $Z_{\rm g}$, $\eta$
or $D/M$, relaxes the sensitivity limit required for detection to similar order of magnitude. We have verified that similar trends are obtained if adopting
survey areas of 0.1 and $10 \ {\rm deg}^2$. Thus, observationally determining the redshift horizon for dust-emitting galaxies could
constrain key properties of primeval galaxy populations.

Finally, we would like to point to encouraging discoveries in recent years of very rare massive sources in wide and shallow surveys, such as those expected to be carried out by near-future FIR telescopes. 
These objects seem to host massive reservoirs of dust in the early universe, while showing many similarities with local galaxies.
\citet{riechers2013}, for example, reported the identification of a rare massive galaxy at
$z\approx6.3$, while searching 21 ${\rm deg}^2$ of the {\em Herschel}/SPIRE data of the
HerMES blank field survey at 250-500 $\mu {\rm m}$. For the brightest
detected target, these authors estimated a FIR luminosity, 
an SFR ($\approx2.9 \times 10^3 \ {\rm M}_{\sun} \ {\rm yr}^{-1}$), 
and gas ($\approx10^{11} \ {\rm M}_{\sun}$) 
and dust ($\approx1.3\times10^{9} \ {\rm M}_{\sun}$)
masses 15-20 times those 
of the prototypical local ultra-luminous starburst Arp 220. They also reported a gas-to-dust
ratio of $\sim80$ and a highly enriched ISM for this object.
More recently, \citet{marrone2018} presented observations of a FIR luminous source at $z\approx6.9$, 
which seems to be composed of two extremely massive 
star-forming galaxies.
The larger of these systems shows an SFR of $\approx2.9 \times 10^3 \ {\rm M}_{\sun} {\rm yr^{-1}}$,
a gas mass of $\approx2.7\times10^{11} \ {\rm M}_{\sun}$ and a dust mass of $\approx2.5\times10^{9} \ {\rm M}_{\sun}$, being more massive than any other known galaxy at $z>6$.  In addition, its companion
shows a stellar mass of $\approx3.5\times10^{10} \ {\rm M}_{\sun}$ and an SFR of
$\approx5.4 \times 10^2 \ {\rm M}_{\sun} {\rm yr^{-1}}$, but with an order of magnitude less
gas and dust.  These authors suggested that these objects could be hosted by a dark-matter halo
with a mass of more than $4\times10^{11} \ {\rm M}_{\sun}$, which would be among the rarest
systems at these early times.  Also recently, \citet{spilker2018} detected a massive molecular outflow from a gravitationally lensed DSFG, 
which likely resides in a dark matter halo of $\sim 10^{12} \ {\rm M}_{\sun}$.
The DSFG studied by \citet{spilker2018} was discovered in the 2500~deg$^{2}$ South Pole Telescope Survey
on the basis of its thermal dust emission and contains $\sim 1.2\times 10^{10} \ {\rm M}_{\sun}$ of molecular gas,
$\sim 1.2\times 10^8 \ {\rm M}_{\sun}$ of dust and an SFR of $\sim 790 \ {\rm M}_{\sun} \ {\rm yr}^{-1}$.  According to earlier
ALMA observations, this source is located at $z\approx5.3$.
Although challenging, the number of detected 
rare massive objects at very high $z$, with extremely high dust-to-gas ratios and high metallicities, will probably
increase during future surveys with next-generation FIR telescopes, as our results indicate.

\section{Comparison with ALMA and NOEMA}
\label{sec:comparison}

\begin{figure*}
        \begin{center}
\plottwo{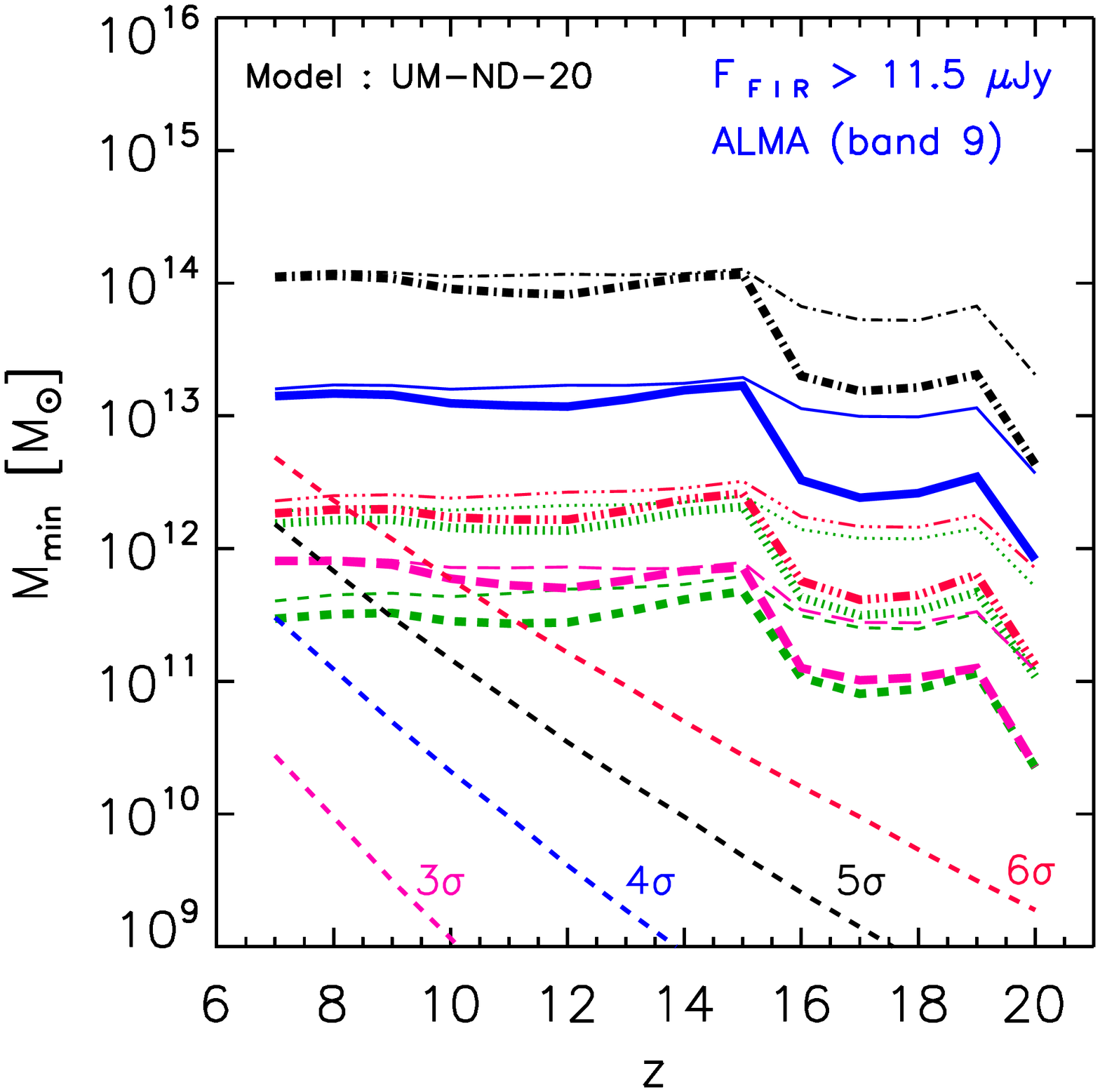}{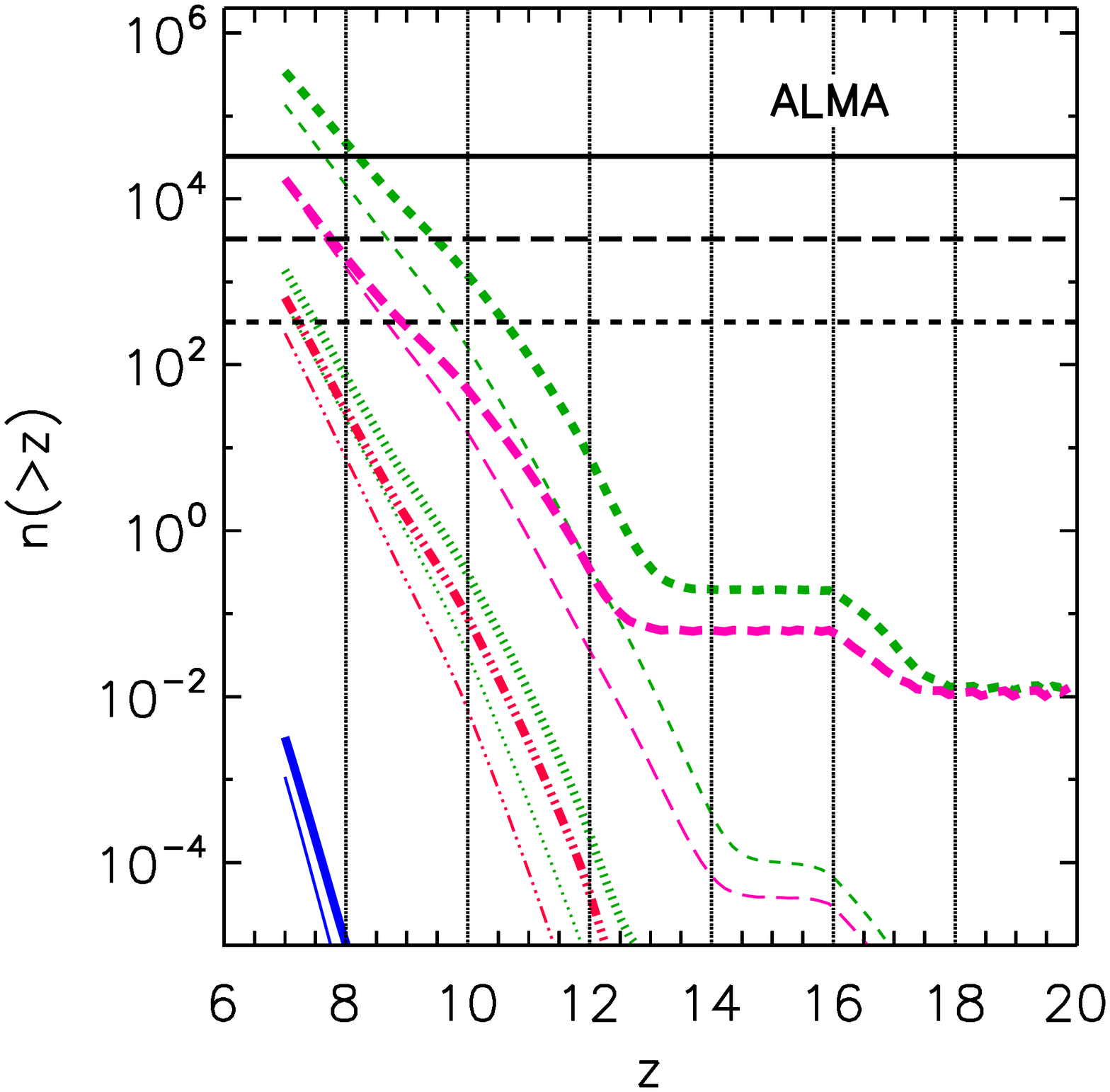}
        \end{center}
    \caption{
Similar to Fig. \ref{fig:variations_Mmin_n}, but for ALMA band 9 (602\-- 720 GHz), assuming
an array configuration of 50 (maximum value) 12-m antennae.
A band-averaged sensitivity of $11.5 \ {\rm \, \mu Jy}$ ($10^6 {\rm \,s}$)
has been assumed (sensitivity as a function of frequency was obtained from the ALMA on-line
sensitivity calculator).
        }
    \label{fig:variations_Mmin_n_ALMA}
\end{figure*}

\begin{figure*}
        \begin{center}
\plottwo{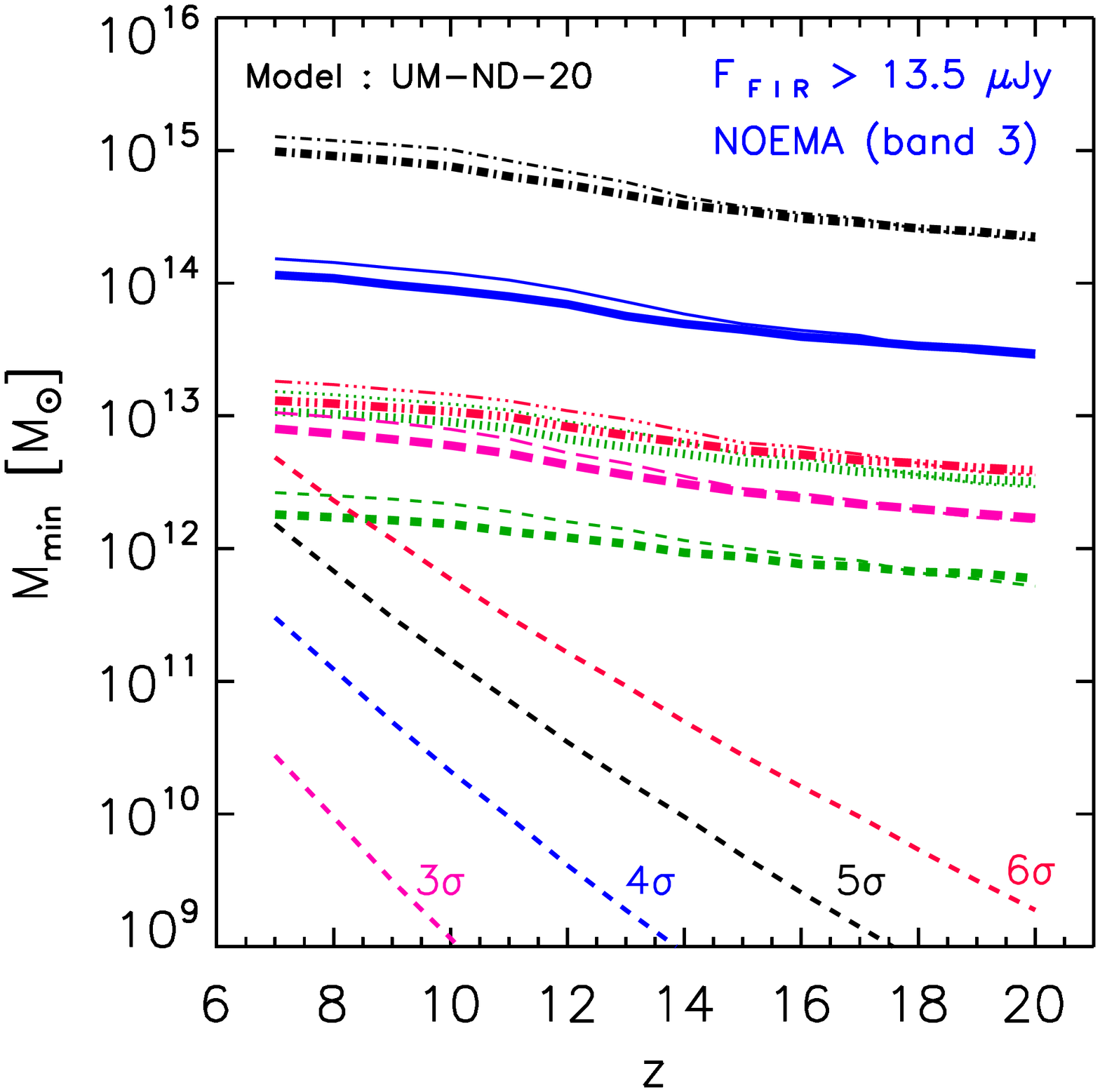}{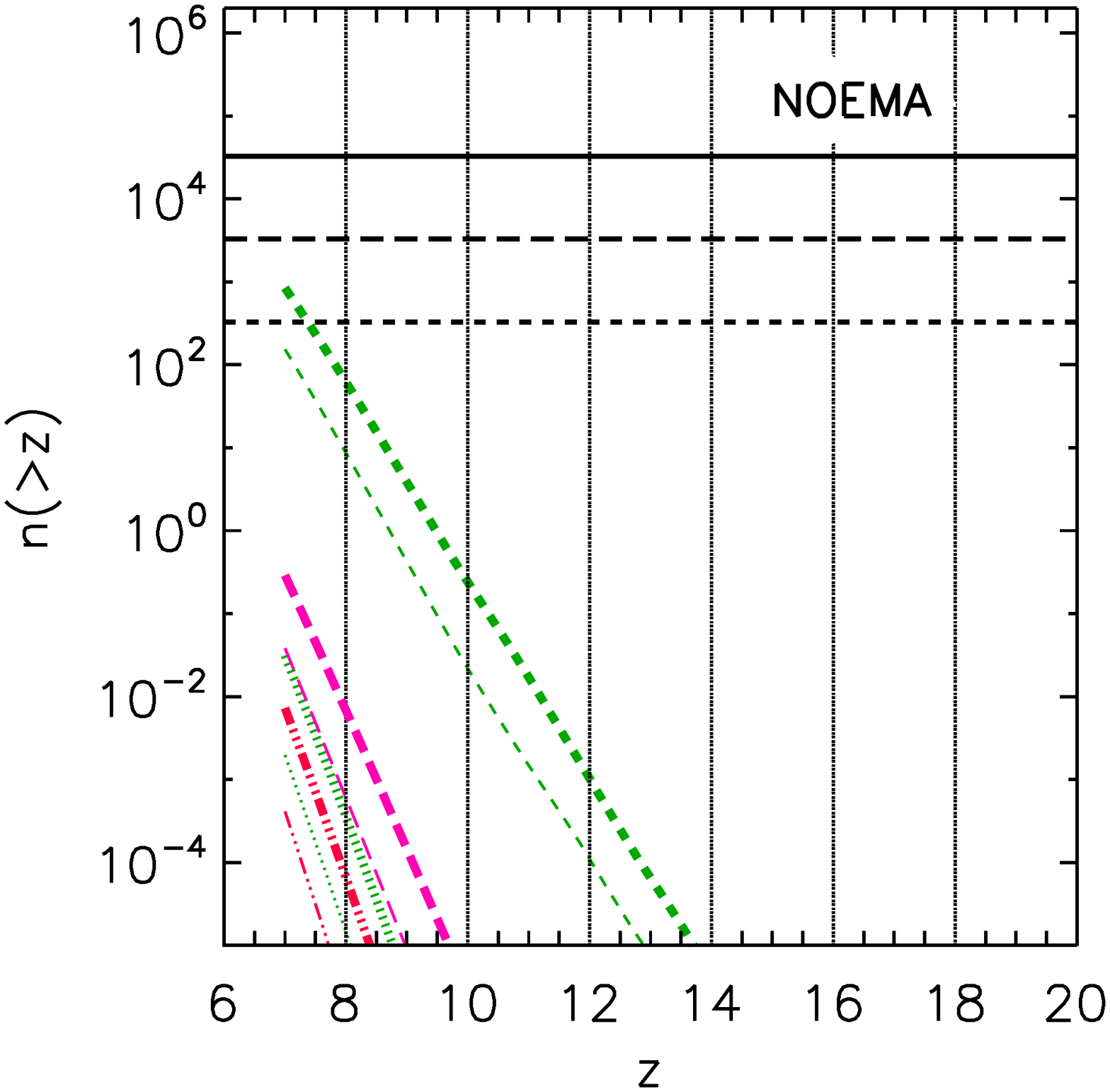}
        \end{center}
    \caption{
	    Similar to Fig. \ref{fig:variations_Mmin_n}, but for NOEMA band 3 (200\-- 276 GHz).
A band-averaged sensitivity of $13.5 \ {\rm \, \mu Jy}$ ($10^6 {\rm \,s}$)
has been assumed.
        }
    \label{fig:variations_Mmin_n_NOEMA}
\end{figure*}

For evaluating the potential improved capabilities of next generation FIR space telescopes to detect the first galaxies, we carry out an approximate comparison
with current ground-based millimeter facilities, specifically ALMA and NOEMA. Furthermore, in Section~\ref{sec:JWST}, we analyse the potential synergy between
such FIR telescopes and the {\it JWST}. In all cases, we consider the capability of each instrument to detect primeval sources for an ultra-deep exposure time of $10^6$\,s.

In the case of ALMA, we assume an optimum array configuration for highest sensitivity, obtained by using fifty 12\,m antennae.
We consider observations within the frequency range 602-720 GHz (ALMA, band 9).
By using the ALMA on-line sensitivity
calculator\footnote{http://almascience.eso.org/proposing/sensitivity-calculator}
and considering a time integration of $10^6$\,s, we derive an average sensitivity of
$11.5 \ {\rm \, \mu Jy}$ for ALMA band 9.
For 12\,m antennae, an average field-of-view (FoV) of $\approx 0.02 \ {\rm arcmin^2}$
is obtained for the latter band.
In the case of NOEMA, we adopt the standard array configuration of eight 15\,m antennae.
We consider observations within the frequency range 200-276 GHz (NOEMA, band 3).
For an exposure of 10\,hr, NOEMA sensitivity
is $\approx 70 \ {\rm \mu Jy}$ for band 3.  Thus,
for an integration time of $10^6$\,s,
we estimate a sensitivity of $\approx 13.5 \ {\rm \mu Jy}$, assuming a scaling with the square root of the exposure time.
Assuming these conditions, an average FoV of $\approx 0.09 \ {\rm arcmin^2}$ is obtained for NOEMA.

By using the aforementioned parameters for ALMA and NOEMA, 
we performed a similar procedure to that described in Sec.~\ref{sec:methodology}
in order to obtain $M_{\rm min}$ and $n$ as a function of $z$.
$M_{\rm min}$ is defined as the lowest $M_{\rm vir}$ such that
$F_{\rm FIR} \ge$ $11.5 \ {\rm \, \mu Jy}$ and $13.5 \ {\rm \, \mu Jy}$, for ALMA and NOEMA,
in their respective wavelength bands.
Results for ALMA can be appreciated in Fig.~\ref{fig:variations_Mmin_n_ALMA}, while
Fig.~\ref{fig:variations_Mmin_n_NOEMA} shows results for NOEMA.
In the case of ALMA (NOEMA), $M_{\rm min} \ga 10^{11} \ {\rm M}_{\sun}$ 
($M_{\rm min} \ga 10^{12} \ {\rm M}_{\sun}$) at $z\ga 7$, even when assuming extremely high
values of $D/M$, $Z_{\rm g}$ and $\eta$. These systems are very massive and rare, according 
to the statistics of the $\nu - \sigma$ peaks, rendering their detection challenging.
We also note that $M_{\rm min}$-$z$ relations are smoother for NOEMA than for ALMA; this is due
to the lower frequencies associated to NOEMA band 3, which is not affected by prominent 
spectral features (see Fig.~\ref{fig:Lnu_fnu_vs_nu}). 

In the right-hand panels of Fig.~\ref{fig:variations_Mmin_n_ALMA} and Fig.~\ref{fig:variations_Mmin_n_NOEMA},
we can see the number density of sources, $n(>z)$, as a function of $z$.
The horizontal lines depict the regions corresponding to the redshift horizon, where $n(>z_{\rm lim}) = 1 / {\Delta}{\Omega}$, considering sky regions of
${\Delta}{\Omega} = 0.1$ (solid line), $1.0$ (long-dashed line) and $10 \ {\rm deg}^2$ (dashed line), respectively.
In the case of ALMA, models with $D/M \ga 0.06$, $Z_{\rm g} \ga 5\times 10^{-2} \ Z_{\sun}$ and $\eta = 0.1$
predict $n(>z) \ga 1/10 \ {\rm deg}^2$ at $z \la 10$, while, for NOEMA, only the model that assumes $\eta = 0.1$
reaches such values at $z\approx7$.  However, considering that the FoV of ALMA (NOEMA) is 
$\approx 0.02 \ {\rm arcmin^2}$ ($\approx 0.09 \ {\rm arcmin^2}$), it is clear that the probability of detecting
sources within one FoV is negligible; even allowing for mosaicing, any detection will be very challenging.

Our results suggest that typical first galaxies hosting primordial dust grains are generally too faint for detection with current ground-based millimeter facilities. The development of next-generation telescopes is thus crucial to enhance the prospects of reaching these elusive sources.

\begin{figure}
        \begin{center}
\plotone{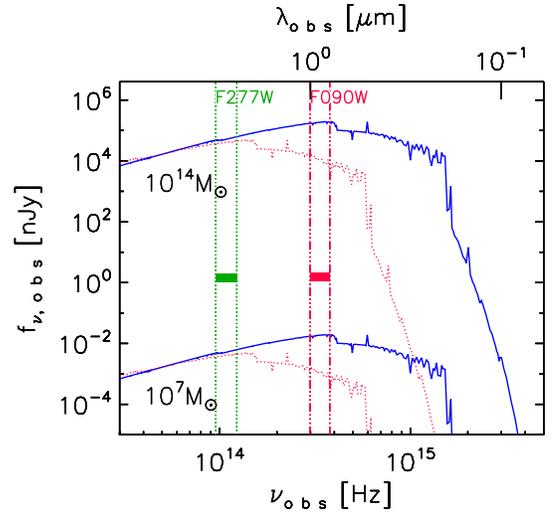}
        \end{center}
    \caption{
Observed fluxes from stellar clusters hosted by halos of
$M_{\rm vir}= 10^{14} \ {\rm M}_{\sun}$ ($M_{*} \approx 1.8\times 10^{11} \ {\rm M}_{\sun}$, upper curves) and $M_{\rm vir}= 10^{7} \ {\rm M}_{\sun}$
($M_{*} \approx 1.8\times 10^{4} \ {\rm M}_{\sun}$, lower curves) at $z=20$ (red dotted curves) and $z=7$ (blue solid curves). Vertical lines
indicate the wavelength ranges associated with filters F277W and F090W
of the NIRCam instrument on {\it JWST}.
The horizontal lines within each filter range depict the associated
sensitivity for an exposure time of $10^6$\,s. This sensitivity together with a FoV of 10
${\rm arcmin}^2$ were used to estimate the critical mass $M_{\rm min}$ and the number
of detected sources $N$, as a function of $z$.
        }
    \label{fig:stellar_spectra}
\end{figure}

\section{Comparison with the JWST}
\label{sec:JWST}
In order to evaluate the potential synergy between next-generation FIR telescopes
and the {\it JWST},
we carry out a rough comparison between our predictions for a future telescope at FIR wavelengths
with those derived for the {\it JWST} at NIR wavelengths.
When comparing with the {\it JWST}, as our model does not include the physical prescriptions required to model the observed NIR radiation self-consistently, we obtain a rough estimate of upper limits for the NIR observed fluxes by simply analysing the stellar emission of our primordial systems.

Our dust model was developed with the aim of studying the FIR/sub-mm
signatures associated with primordial galaxies and, hence, it is focused on the physics of dust emission.
As discussed in \citet{derossi2017}, the model does not include the explicit physical
prescriptions required to model observed spectra in NIR wavebands, such as nebular
emission or the reprocessing of Lyman-$\alpha$ photons in the IGM.
Thus, to obtain a rough estimate of the NIR luminosities of our sources, we evaluate the fluxes associated with the stellar
component of our model galaxies, as a function of $z$ and $M_{\rm vir}$. For simplicity, we assume negligible extinction, so that the predicted fluxes represent upper limits.

According to some high-$z$ observations \citep[e.g.,][and references therein]{fudamoto2017}, 
typical observed IR-to-UV ratios range between $\sim 0.1$ and $\sim 100$
at $M_* \sim 10^9 - 10^{12}\ {\rm M_{\sun}}$.\footnote{We emphasize that these values correspond to upper limits in many cases.} 
At similar masses and for our fiducial model parameters (see Sec.~\ref{sec:methodology}), 
we obtain IR-to-UV ratios\footnote{We define the IR-to-UV ratio as
$L_{\rm IR}/ L_{\rm UV}$, where $L_{\rm IR}$ corresponds to rest-frame bolometric dust luminosity
at $\lambda = 8 - 1000 \ \mu{\rm m}$, and $L_{\rm UV}$  is the rest-frame stellar luminosity at 
$1600 \ {\rm \AA}$ with $L_{\rm UV} = \nu(1600 \ {\rm \AA}) L_{\nu}(1600 \ {\rm \AA})$.}
ranging between $\sim 0.001$ and $\sim 1$, depending on mass, $z$ and dust model.
As we are modelling the first galaxies hosting dust grains, our dust densities are very low. It is thus expected to obtain low IR-to-UV ratios, compared to galaxies in more advanced evolutionary stages, such as those currently observed.  
We note, however, that an increase in the dust-to-metal
ratios, gas metallicities and/or star formation efficiencies of our model sources leads to an increase in the IR-to-UV ratios, reaching 
values similar to those currently observed at high $z$ for the extreme cases considered in Sec.~\ref{sec:parameter_variations}.

\begin{figure*}
        \begin{center}
\plottwo{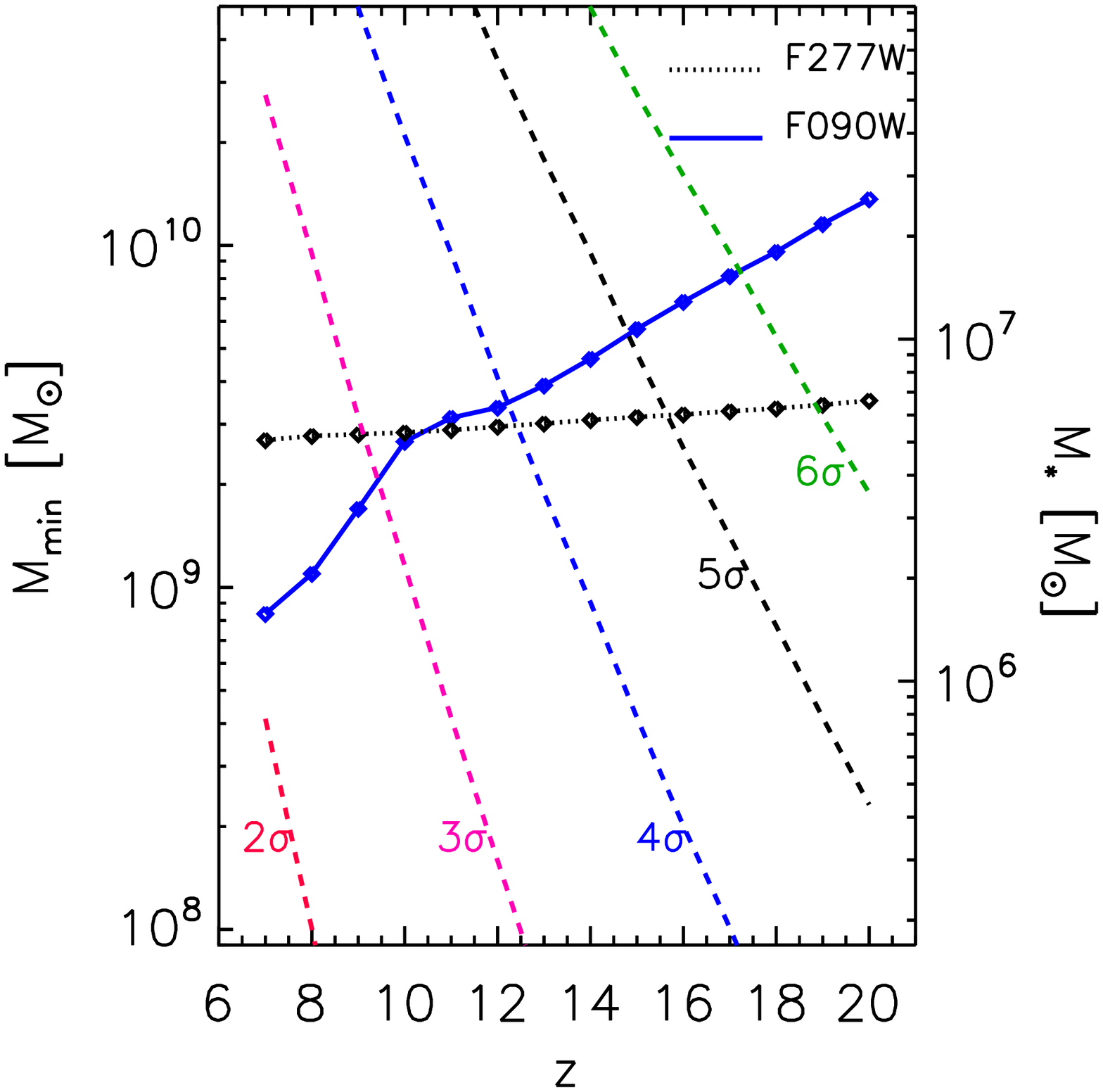}{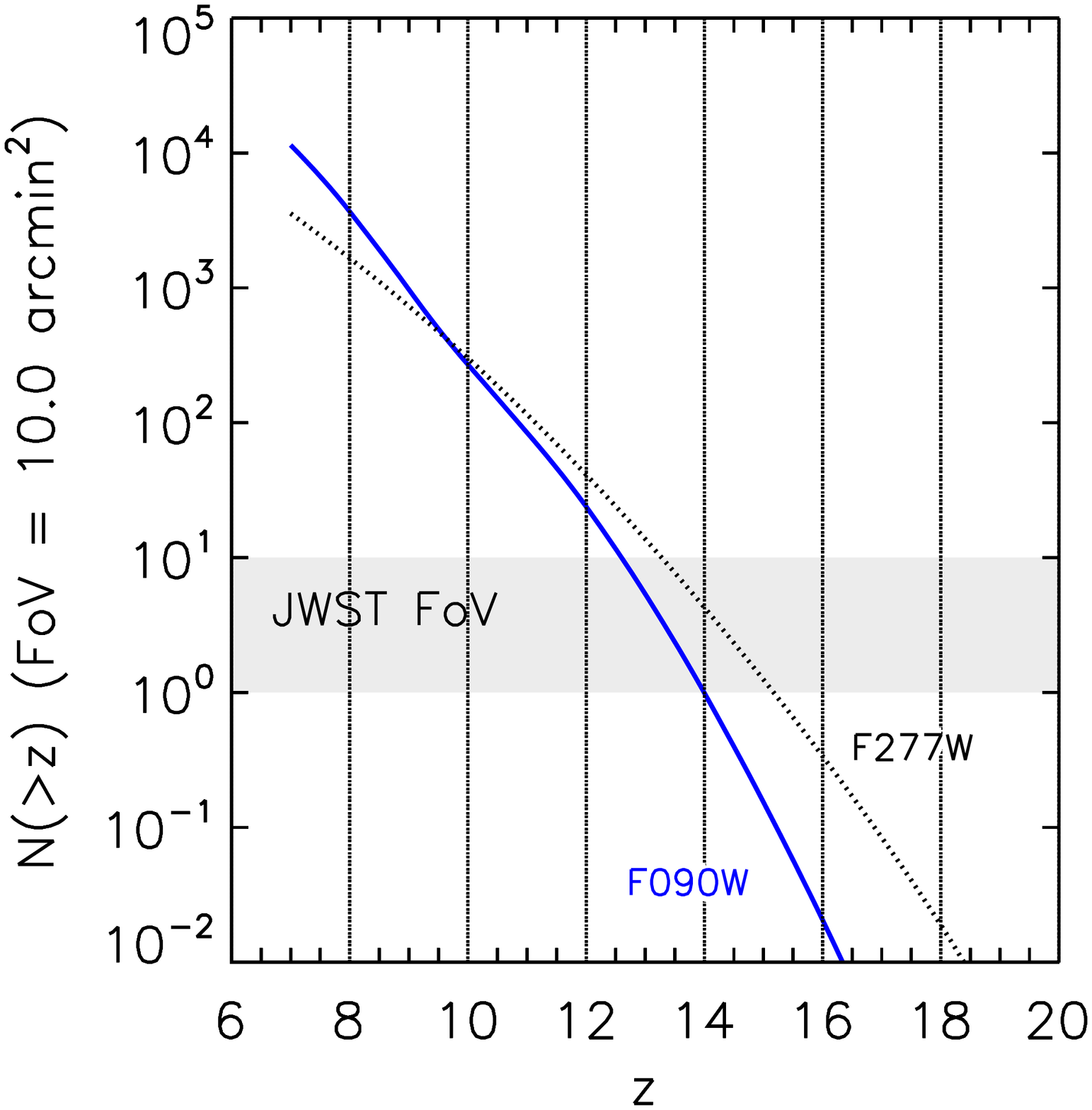}
        \end{center}
    \caption{
Deep surveys with the {\it JWST}. $M_{\rm min}$ (left-hand panel) and $N$ (right-hand panel) as a function of $z$, obtained when using
the NIRCam F277W and F090W filters.
	In the left-hand panel, the right vertical axis shows $M_*$ as a reference.
In the right-hand panel, a ${\rm FoV} = 10 \ {\rm arcmin}^2$ has been adopted.
Although this is a rough estimate of the possible contribution of the stellar sources to
the NIR radiation, we see that the number of detected sources per FoV reaches 1-10
at $z\sim15-13$.  Thus, observations in the NIR could help to guide possible surveys
with next-generation FIR telescopes.
        }
    \label{fig:variations_JWST}
\end{figure*}

In Fig.~\ref{fig:stellar_spectra}, we show the observed specific fluxes associated with
stellar clusters hosted by systems of $M_{\rm vir} = 10^7 \ {\rm M}_{\sun}$ ($M_{*} \approx 1.8\times 10^{4} \ {\rm M}_{\sun}$, lower
curves) and $M_{\rm vir} = 10^{14} \ {\rm M}_{\sun}$ ($M_{*} \approx 1.8\times 10^{11} \ {\rm M}_{\sun}$, upper curves) at $z=7$ (blue solid
curves) and $z=20$ (red dotted curves). We evaluate the probability of detecting these systems with the Near Infrared Camera (NIRCam) F277W and F090W filters on board the {\it JWST}. The wavelength ranges and sensitivity levels (for an exposure time of $10^6$\,s) 
associated with the filters are indicated in Fig.~\ref{fig:stellar_spectra}
by using vertical and horizontal lines, respectively.
For estimating $M_{\rm min}$ and $N$, we follow a similar approach to that described
in Sec.~\ref{sec:methodology}.  We calculate the average NIR flux ($F_{\rm NIR}$) within each filter 
by using an expression similar to Equ.~\ref{eq:average_flux}, and define $M_{\rm min}$ as the lower virial mass, such that $F_{\rm NIR}$ is above the sensitivity level within each filter.
In order to estimate $N$, we assume a FoV of $\Delta \Omega =$ 10 ${\rm arcmin}^2$ for the NIRCam instrument.

Predictions for the {\it JWST} are shown in Fig.~\ref{fig:variations_JWST}.
We can see that $M_{\rm min} \ga 10^9 \ {\rm M}_{\sun}$ for both the F277W and
F090W NIRCam filters, along the entire redshift interval considered.
In particular, $M_{\rm min} \sim 3 \times 10^{9} \ {\rm M}_{\sun}$ 
for  filter F277W, not exhibiting significant variations with $z$, as expected from Fig.~\ref{fig:stellar_spectra}, since the specific flux
within this filter barely evolves.  On the other hand,
$M_{\rm min}$ decreases from $10^{10} \ {\rm M}_{\sun}$ at $z\approx20$ to $10^{9} \ {\rm M}_{\sun}$
at $z\approx7$ in the case of filter F090W, because
of the increase of the specific flux within its associated wavelength range as $z$ decreases.
For our default model parameters, similar $M_{\rm min}$ could be achieved in the FIR for sensitivities
$S\sim 1 \ {\rm nJy}$ (Fig~\ref{fig:Ffir_vs_M}), which are probably too low to be reached in the
next decades. However, as previously discussed, galaxy populations with higher metallicities, 
dust fractions and/or star formation efficiencies could be detected at the FIR for higher sensitivity values.

As can be seen in the right-hand panel of Fig.~\ref{fig:variations_JWST}, the redshift horizon for the {\it JWST} is reached at $z\approx15$ for filters F090W and 
F277W. These findings suggest that upcoming observations in the NIR with the {\it JWST} 
could help to guide possible later surveys in the FIR with the next-generation telescopes.
The synergy between such facilities could be crucial to elucidate the first galaxies at these early times.
However, it is worth highlighting again that the 
comparisons presented in this section 
should be taken with a grain of salt, as our model is not explicitly designed to model the observed NIR radiation.

\section{Summary and Conclusions}

In this study, we explore the prospects of detecting the dust 
continuum emission from first galaxy populations with next-generation FIR telescopes. Specifically,
we apply the dust model developed by \citet{derossi2017} to estimate
the least massive FIR/sub-mm sources that such facilities would be able to detect.
We also predict the redshift horizons to detect at least one source within potential survey regions. 
As the design concepts for future FIR telescopes are evolving,
we focus on a generic instrument, exploring a wide range of plausible
sensitivities and survey areas. Our goal is to provide a baseline of exploration
for future FIR campaigns, applicable to different sets of instrument parameters.
Our main results can be summarized as follows:

\begin{itemize}
\item 
For systems of similar masses, higher dust temperatures are obtained at higher $z$
(Fig. \ref{fig:Td_vs_r}).  
This is caused by the increased intensity of the CMB radiation
at earlier epochs. In addition, at higher $z$, systems of a given mass are more concentrated and 
exhibit higher central dust densities, which contribute to more efficient heating by the 
central stellar cluster.
\item Because of the increase of dust temperature at higher $z$, galaxies of similar masses
tend to be brighter at FIR/sub-mm wavelengths as $z$ increases (Fig.\,\ref{fig:Lnu_fnu_vs_nu} and
\ref{fig:Ffir_vs_M}). This effect leads to an extremely
negative K-correction for these FIR/sub-mm sources, compensating and even reversing the decrease
of the observed flux with the distance to the source (Fig.~\ref{fig:Ffir_vs_z}).
\item As a consequence of the strong negative K-correction,
the threshold virial mass that a generic FIR telescope would be able to detect during a survey campaign reaches a minimum
at $z>12$.  For an instrument sensitivity $S=1 \ {\mu}{\rm Jy}$, 
$M_{\rm min} \approx 1-2 \times 10^{12} \ {\rm M}_{\sun}$ for the standard dust size
distribution, and $M_{\rm min} \approx 3-4 \times 10^{11} \ {\rm M}_{\sun}$ for the shock size
distribution (Fig.\,\ref{fig:Mmin_vs_z}).
Interestingly, we also found that at $7<z<12$ and for some dust models, the threshold mass increases towards lower $z$.  This latter unexpected result is driven by different spectral features that enter
the assumed photometric filter as $z$ increases.
Different sensitivity values lead roughly to similar trends but with an increase of the absolute
normalization of $M_{\rm min}$ for higher $S$ (Fig.\,\ref{fig:Ffir_vs_M}).
\item Although the strong negative K-correction associated with the first galaxies would contribute to
their detection, these galaxies are located in very rare overdensities.
Thus, they will be difficult to detect in blind surveys.  Gravitational lensing might boost detectability, to be explored in future work.
\item Variations from our fiducial values for dust-to-metal ratio (0.005 -default-, 0.06, 0.4)
gas metallicity ($5\times10^{-4}$, $5\times10^{-3}$ -default-, $5\times10^{-2} \ {\rm Z}_{\sun}$), 
or star formation efficiency (0.01 -default-, 0.1), can change the minimum halo mass needed for detection by more than two orders of magnitude (Fig.\,\ref{fig:variations_Mmin_n}, left-hand panel).
\item The probability of detecting one individual {\em typical} first galaxy
during a potential FIR survey depends mainly on the instrument sensitivity ($S$)
and, secondly, on the survey area ($\Delta \Omega$) (Fig.~\ref{fig:n_vs_z}).
In the case of the shock size distribution, the redshift horizon
for $S\sim 1 \ {\mu}{\rm Jy}$ is reached at $z\sim 7-10$, with
the exact value depending on ${\Delta}{\Omega}$ and dust chemical composition.
In the case of the standard size distribution, the redshift horizon
for $S\sim 1 \ {\mu}{\rm Jy}$ only reaches $z\sim7$ for
${\Delta}{\Omega}=10\ {\rm deg}^2$ and for some particular dust chemistries.
For instrument sensitivities $S \la 0.1 \ {\mu}{\rm Jy}$ and
${\Delta}{\Omega} \ga 0.1 \ {\rm deg}^2$, all dust models
predict a redshift horizon above $z\approx7$.
\item Interestingly, for all dust models, 
the redshift horizon decreases roughly linearly with ${\log}_{10}(S)$ at $z>12$,
exhibiting a similar slope of $\approx 4$ (Fig.~\ref{fig:redshift_horizon}).  At $z<12$, some models predict departures from such linear relation as different spectral features enter the adopted FIR band.
\item The FIR redshift horizon strongly depends on the properties of primordial dust 
and its host galaxy population.
Considering that the nature of systems hosting primordial dust is very uncertain,
with some recent studies suggesting high dust fractions in massive galaxies already at early epochs,
we explore the effects of varying some critical galaxy properties such as gas metallicity,
dust fractions and the star formation efficiency (Fig.~\ref{fig:variations_Mmin_n} and
\ref{fig:variations_redshift_horizon}).
For $S=1 \ {\mu}{\rm Jy}$ and assuming a UM-ND-20 dust chemistry, metallicities
$Z_{\rm g} \ga 5\times10^{-2} \ {\rm Z}_{\sun}$
or dust fractions $D/M\ga0.06$ are required
to reach the redshift horizon at $z > 7$ for $\Delta \Omega=0.1-10 \ {\rm deg}^2$.
However, these limits could move
towards lower values for different dust chemical compositions.
For galaxies below such metallicity and dust fraction limits,
gravitational lensing would be required to detect these fainter systems if sensitivity limits cannot
be boosted. 
\item In general, an increase by one order of magnitude of $Z_{\rm g}$, $\eta$ or $D/M$, 
similarly relaxes the sensitivity limit
required for the galaxy population to reach the
redshift horizon at a given $z$ (Fig.~\ref{fig:variations_redshift_horizon}).
\item  Typical first galaxies hosting primordial dust grains seem to be too faint to be
	detected by current ground-based millimeter facilities, such as ALMA and NOEMA 
	(see Fig.~\ref{fig:variations_Mmin_n_ALMA} and \ref{fig:variations_Mmin_n_NOEMA}). Next-generation
	FIR telescopes are thus likely to play an important role in searching for these elusive sources.
\item  Although our model is not specifically designed to describe the
NIR radiation from primeval galaxies, a rough estimate suggests a redshift horizon of $z\sim15$ for the {\it JWST} NIRCam instrument (Fig.\,\ref{fig:variations_JWST}). 
Hence, observations with the {\it JWST} in the near future could guide follow-up surveys with next-generation 
FIR telescopes.
\end{itemize}

We conclude that the detectability of the dust continuum emission from first galaxies as 
FIR/sub-mm sources is boosted by an extremely negative K-correction. 
However, as these systems would reside in rare high-density
peaks, their detection in blind surveys will be challenging.
Such programs would require the development of techniques to address confusion noise. 
In particular, the detection of extremely faint sources could likely require flux amplification by gravitational lensing.
Finally, taking into account the sensitivity of the FIR redshift horizon to the assumed
dust properties, measurements of FIR/sub-mm source densities at $z>7$
would shed light on the nature of dust at the very dawn of star formation.

\acknowledgments

We thank the anonymous referee for constructive comments, which helped to improve this paper.
We also thank Alexander Ji for providing tabulated dust opacities for the different
dust models used here.
This work makes use of the Yggdrasil code \citep{zackrisson2011}, which adopts
Starburst99 SSP models, based on Padova-AGB tracks \citep{leitherer1999, vazquez2005}
for Population~II stars. 
VB acknowledges support from NSF grant AST-1413501.
MEDR is grateful to PICT-2015-3125 of ANPCyT (Argentina) and also to
Mar\'{\i}a Sanz and Guadalupe Lucia for their help and support.

\end{document}